\PassOptionsToPackage{dvipsnames}{xcolor}
\documentclass[floatfix,aip,reprint,superscriptaddress,jcp]{revtex4-1}

\usepackage{preamble}

\begin{document}

\title{Spectral Density Modulation and Universal Markovian Closure of Fermionic Environments}

\affiliation{\UnimiFisica}
\affiliation{\UlmPhysik}
\affiliation{\INFN}

\author{Davide Ferracin}
\affiliation{\UnimiFisica}

\author{Andrea Smirne}
\affiliation{\UnimiFisica}
\affiliation{\INFN}

\author{Susana~F.~Huelga}
\affiliation{\UlmPhysik}

\author{Martin~B.~Plenio}
\affiliation{\UlmPhysik}

\author{Dario Tamascelli}
\affiliation{\UnimiFisica}
\affiliation{\UlmPhysik}

\begin{abstract}
The combination of chain-mapping and tensor-network techniques provides a powerful tool for the numerically exact simulation of open quantum systems interacting with structured environments.
However, these methods suffer from a quadratic scaling with the physical simulation time, and therefore they become challenging in the presence of multiple environments.
This is particularly 
true when fermionic environments, well-known to be highly correlated, are considered.
In this work we first illustrate how a thermo-chemical modulation of the spectral density allows replacing the original fermionic environments with equivalent, but simpler, ones.
Moreover, we show how this procedure reduces the number of chains needed to model multiple environments.
We then provide a derivation of the fermionic Markovian closure construction, consisting of a small collection of damped fermionic modes undergoing a Lindblad-type dynamics and mimicking a continuum of bath modes.
We describe, in particular, how the use of the Markovian closure allows for a polynomial reduction of the time complexity of chain-mapping based algorithms when long-time dynamics are needed.
\end{abstract}

\maketitle

\noindent
\emph{\small
  This article may be downloaded for personal use only. Any other use requires prior permission of the author and AIP Publishing. This article appeared in
  \begin{quoting}
    Davide Ferracin, Andrea Smirne, Susana F.~Huelga, Martin B.~Plenio, Dario Tamascelli. \emph{Spectral density modulation and universal Markovian closure of fermionic environments}. J.~Chem.~Phys. 7 November 2024, 161~(17)
  \end{quoting}
  and may be found at \href{https://doi.org/10.1063/5.0226723}{doi.org/10.1063/5.0226723}.
}

\noindent
\emph{\small
  Copyright 2024 The Authors.
  This article is distributed under a \href{https://creativecommons.org/licenses/by/4.0/}{Creative Commons Attribution 4.0 International (CC BY) License}.%
}

\section{Introduction} \label{sec:intro}
The theory of open quantum systems~(OQS) provides a framework to study quantum systems when their interaction with the surrounding environment---be it desired or undesired---is not negligible and leads to the emergence of irreversible and noisy processes~\cite{breuer02,rivas12,vacchini24}.
In certain situations, it is possible to model such an interaction as resulting in white noise, amenable to an effective description in terms of a Gorini-Kossakowski-Sudarshan-Lindblad~(GKSL) master equation~\cite{Lindblad1976, Gorini1976}.
In general, however, the system and the environment mutually influence each other's evolution on timescales relevant to the process under study; this gives rise to non-Markovian effects that cannot be captured by the GKSL master equation: details on the past state of the system, on its many body environment and on the correlations between them must be retained.
Solid state implementations of qubits, such as quantum dots~\cite{VanDerViel:kondo_effect_unitary_limit,Sasaki:kondo_integer_spin_quantum_dot}, molecular transistors~\cite{Etchegoin:molecular_transistors}, unconventional superconductors~\cite{Fritz:kondo_screening_unconventional_superconductors}, and molecular magnets~\cite{Romeike:kondo_transport_spectroscopy}, quantum thermal machines~\cite{Strasberg2016, Pta2022}, quantum sensing and metrology protocols~\cite{Chin2012, Smirne2016} ,
energy-charge conversion and exciton transport in solid-state devices~\cite{Gelvez-Rueda:2020aa, alvertis2020}, or synthetic or biological light-harvesting complexes~\cite{Lim:2015aa, Romero:2014aa} are typical instances in which deviations from a Markovian evolution can play a significant role.

Due to the large number of environmental degrees of freedom affecting the reduced dynamics of the OQS under
investigation, the simulation of even simple OQSs represents a most challenging computational task.
Beside exploiting methods originally developed for the simulation of strongly correlated quantum systems, such as the numerical renormalization group in the basis of scattering states~\cite{Anders:steady_state_currents_nanodevices, Schmitt:comparison}, flow equations~\cite{Wegner:flow_equations, Kehrein:flow_equation_many_particle}, the time-dependent density-matrix renormalization group~\cite{Schollwoeck:dmrg, Schollwoeck2011:dmrg_in_the_age_of_mps}, multilayer multiconfiguration time-dependent Hartree (ML-MCTDH)~\cite{Wang:multilayer_multiconfiguration_time_dependent_hartree},  and continuous-time quantum Monte Carlo~\cite{Cohen:taming_dynamical_sign_problem, Dong:quantum_monte_carlo_solution, Ridley:full_counting_statistics_nonequilibrium_anderson}, a number of algorithms specifically designed for the simulation of OQSs dynamics have been proposed in recent years.
Hierarchical Equation of Motion (HEOM)~\cite{Tanimura2020:heom}, Dissipation-Assisted Matrix Product Factorization (DAMPF)~\cite{Somoza19:dampf, Somoza22:dampf}, Time Evolving Matrix Product Operators (TEMPO)~\cite{Strathearn2018:tempo} and Automated Compressed Environment (ACE)~\cite{Cygorek2022:automated_compressed_environments} are representative of this class of OQS simulation methods.

Chain-mapping techniques combined with matrix-product representations of quantum states represent a powerful tool for the non-perturbative simulation of open quantum systems interacting with structured environments.
Chain-mapping consists in a reshaping of the environment into a one dimensional lattice of modes with nearest-neighbor interactions, a structure that is, in general, very well suited for DMRG-related techniques~\cite{wolf14}.
While different chain mappings have been introduced, the Time Evolving Density Operator with Orthogonal Polynomials (TEDOPA)~\cite{prior10, chin10, woods14} provides the most accurate discretization of continuous spectral densities~\cite{devega15discr}.
The TEDOPA algorithm has found application to the study of a variety of open quantum systems strongly interacting with highly structured environments~\cite{chin13, Caycedo-Soler:2022aa, Lacroix2024}.
After the mapping, TEDOPA determines the evolution of both the OQS and the environmental degrees of freedom on the same footing.
This is a most remarkable feature of the algorithm paving the way to the investigation of fundamental OQS phenomena, such as the mechanism at the origin of polaron formation in spin-boson models~\cite{Riva2023:polaron_formation_bosonic}.
The 1-dimensional topology, moreover, implicitly defines a light-cone structure that allows systematically determining the effective reduced environment which is relevant to  the system dynamics within any assigned evolution time, leading to a further major reduction of the computational complexity and allowing for the simulation of more complex open quantum systems~\cite{Lorenzoni2024:systematic_coarse_graining}.

One of the main limitations of the original TEDOPA formulation is its quadratic scaling, $O(t^2)$, with the physical simulation time $t$~\cite{Tamascelli2019:ttedopa}.
As a matter of fact, actual computer simulations require a truncation of the semi-infinite chain resulting from the TEDOPA mapping after a finite number of sites.
Such a truncation must be suitably chosen so as not to introduce finite size effects~\cite{Woods2015,woods2016dynamical}.

Longer simulation times, therefore, will require longer chains.
A solution to this problem has been put forward, for the case of bosonic environments, in Ref.~\onlinecite{Nuesseler2022:markovian_closure} with the Markovian Closure~(MC) construct, namely a small collection of damped harmonic modes able to mimic a semi-infinite uniform chain of harmonic oscillators.
The use of the MC allows for a reduction to $O(t)$ of the TEDOPA time complexity and makes it possible to use the algorithm to determine long time OQS dynamics, such as those required for one- and two-dimensional electronic spectroscopy.

Another context where long time dynamics are typically required is the investigation of non-equilibrium steady-state properties (NESS) of open quantum systems interacting with fermionic environments.
Macroscopic fermionic environments play a major role in several areas of condensed matter physics, such as the Kondo effect, explaining the resistance minimum of metallic conductors.
Another key area where fermionic environments play a crucial role is the dynamical mean-field theory (DMFT) approach to strongly correlated materials.
The challenging part of the program is to solve the impurity model; this requires finding the impurity spectral function, which can be accurately determined only if a NESS is reached.
Fermionic environments typically appear also in the study quantum transport, where two fermionic environments (leads) at a certain temperature and chemical potential are connected to some quantum system that acts as a bridge to transport energy and/or particles~\cite{mitchison2018non,wojtowicz2020open,zwolak20}.

The extension of the TEDOPA mapping to fermionic environments has been discussed in detail in Ref.~\onlinecite{Nuesseler2020:ftedopa}.
The analysis presented in Ref.~\onlinecite{Kohn2021:interleaved_mapping}, in particular, reveals a steady and fast build-up of strong correlations between the environmental fermionic modes induced by the interaction with the impurity.
Even by adopting state-of-the art strategies for the mitigation of the effects of the appearance of such correlations, the determination of the OQS dynamics over long times, which is typically required to reach the NESS state, remains a computationally most challenging task.
The possibility of exploiting the MC construct in the fermionic setting would therefore represent a powerful asset.
The definition of a MC mechanism for the case of fermionic structured environments is, however, currently lacking. In this work we fill this gap by introducing a fermionic MC, and illustrate its application and computational impact by means of some relevant case studies.

The manuscript is organized as follows.
We start by defining the general model in \Cref{sec:model}.
We discuss the thermalization of the spectral density in \Cref{sec:fermionic_spectral_density_thermalization} and provide details on the chain-mapping procedure in \Cref{sec:tcsm-tedopa}.
\Cref{sec:fermionic_markovian_closure} is devoted to the derivation of the Fermionic Markovian Closure~(FMC).
In \Cref{sec:numerical_tests} we present the accuracy/performance results.
We conclude the work by discussing other possible applications of the FMC construct.

\section{The model} \label{sec:model}
Consider a general impurity model interacting with continuous fermionic baths (leads).
The evolution of the overall system-bath complex is defined by the Hamiltonian
\begin{align}
    H_{\System*\Environment*}&=
    H\System + H\Environment + H\Interaction,
    \label{eq:fullHam}\\
    H\Environment&=H_{\Environment*,\Left*} + H_{\Environment*,\Right*} =
    \sum_{\alpha\in\{\Left*,\Right*\}} \int_{\Omega_\alpha}\dd \omega\,(\omega-\mu_\alpha) \adj{f_{\alpha,\omega}}f_{\alpha,\omega}\phantomadj,
    \label{eq:hamiltonian_quadratic_env}\\
    H\Interaction&=
  \sum_{\alpha\in\{\Left*,\Right*\}} \sum_{\lambda=1}^{m} \int_{\Omega_\alpha}\dd \omega\, h_{\lambda,\alpha}(\omega) (\adj{A_{\lambda}}f_{\alpha,\omega}\phantomadj+\adj{f_{\alpha,\omega}}A_{\lambda}\phantomadj).
    \label{eq:intHam}
\end{align}
The derivation that follows is essentially independent of the details of the system, which we model here as an $m$-level fermionic system with annihilation and creation operators $d_\lambda$ and $\adj{d}_\lambda$, $\lambda\in\{1,\dotsc,m\}$, satisfying the canonical anticommutation relations (CARs) $\{d_\lambda,\adj{d}_\nu\}=\delta_{\lambda,\nu}$; the system Hamiltonian $H\System$ is an arbitrary function of these operators.
The Hamiltonian $H\Environment$ describes the left ($\Left*$) and right ($\Right*$) leads, i.e.~two continua of (non-interacting) fermionic modes indexed by their frequency $\omega\in\Omega_\alpha$, in the presence of a chemical potential $\mu_\alpha$, with $\alpha \in \{\Left*,\Right*\}$; the operators $f_{\alpha,\omega}$ and $\adj{f}_{\alpha,\omega}$ and obey CARs $\fcomm{f_{\alpha,\omega}\phantomadj,\adj{f_{\alpha',\omega'}}} = \delta_{\alpha,\alpha'}\delta(\omega-\omega')$.

The interaction between the system and the environment is specified by $H\Interaction$, with $A_{\lambda}$ and $f_{\alpha,\omega}$ and $\adj{f}_{\alpha,\omega}$ operators acting on level $\lambda$ of the system and on the left ($\alpha = \Left*$) or right ($\alpha = \Right*$) bath side respectively.
The interaction (hybridization) strength between the system and the fermionic modes is instead given by $h_{\lambda,\alpha}(\omega) = \kappa_{\lambda,\alpha}\sqrt{J_{\alpha}(\omega)}$, where the functions $J_{\alpha}\colon\Omega_\alpha\to\R^+$ and $\alpha\in\{\Left*,\Right*\}$ are spectral densities, while the (real) coefficients $\kappa_{\lambda}$ model possibly different interaction strengths between the environments and the different levels of the system.
We observe that the system-bath complex \cref{eq:fullHam} can describe different relevant fermionic (spinless) open quantum systems such as the resonant level model (RLM) and the interacting resonant level model (IRLM), and can be straightforwardly generalized to the case of spinful fermions (e.g.~SIAM) and multichannel leads.
We point out that if the system Hamiltonian $H\System$ comprises only  terms that are up to quadratic in $d_\lambda$ and $\adj{d}_\lambda$, the model fixed by~\eqref{eq:fullHam} is exactly solvable by means of exact diagonalization~\cite{Jung:2020aa}.

We assume, moreover, that the overall system is initially in a factorized state $\rho_{\System*\Environment*}(0) = \rho\System(0)\rho\Environment(0)$, for an arbitrary $\rho\System(0)$, while $\rho\Environment(0) = \rho_{\Environment*,\Left*}(0)\rho_{\Environment*,\Right*}(0)$ is the factorized state of the left (right) environment in the thermal state \(\rho_{\Environment*,\Left*}(0)\) (\(\rho_{\Environment*,\Right*}(0)\)) for a given chemical potential \(\mu\Left\) (\(\mu\Right\)) and inverse temperature $\beta\Left=1/T\Left$ ($\beta\Right = 1/T\Right$), 
with
\begin{equation} \label{eq:envThermal}
    \rho_{\Environment*,\alpha} =
    \frac{1}{\partitionf_\alpha}\exp(-\beta_{\alpha} H_{\Environment*,\alpha}),
\end{equation}
where \(\partitionf_\alpha=\Tr\exp(-\beta_\alpha H_{\Environment*,\alpha})\), \(\alpha\in\{\Left*,\Right*\}\) is the partition function.
We observe that  correlated system-environment initial states can be prepared starting from this reference state by either an adiabatic evolution (as we will do in \cref{sec:numerical_tests}), or a DMRG search of the (correlated) ground state.
The thermal state at inverse temperature $\beta$ for a single fermion of frequency $\omega$ in the presence of a chemical potential $\mu$ is
\begin{equation} \label{eq:rhobetamu}
    \rho_{\beta,\mu} (\omega)=
    \frac{1}{\partitionf}\exp\bigl(-\beta(\omega-\mu)\adj{f_\omega}f_\omega\phantomadj\bigr).
\end{equation}
The expectation value of the number operator $\avg{\adj{f_\omega}f_\omega\phantomadj}$ satisfies the Fermi-Dirac distribution
\begin{equation} \label{eq:avocc}
   n_{\beta,\mu}(\omega)=
   \Tr\bigl(\rho_{\beta,\mu} (\omega)\adj{f_\omega}f_\omega\phantomadj\bigr)=
   \frac{1}{e^{\beta(\omega-\mu)}+1}.
\end{equation}
A local Hamiltonian term of  the form $(\omega-\mu)\adj{f_\omega}f_\omega\phantomadj$ admits the eigenvalues \(0\) and \(\omega-\mu\), corresponding to the vacuum and filled state that we indicate respectively by $\ket{0}_\omega$ and $\ket{1}_\omega$. Depending on the value of $\mu$, therefore, the ground state of the local Hamiltonian for the mode $\omega$ can be the vacuum or filled state.

\section{Equivalent environments}
\label{sec:fermionic_spectral_density_thermalization}
The model described by \cref{eq:fullHam,eq:hamiltonian_quadratic_env,eq:intHam} satisfies the following three hypotheses: (i)~the free environment Hamiltonian $H\Environment$ is quadratic in \(f_{\alpha,\omega}\) and \(\adj{f_{\alpha,\omega}}\), (ii)~the initial state $\rho\Environment(0)$ of the environment is Gaussian with zero mean of the creation and annihilation operators, i.e.~vanishing first moments, and (iii)~the system-bath interaction is bilinear in $f_{\alpha,\omega}$ and $\adj{f}_{\alpha,\omega}$.
Because of these three properties, the two-time correlation functions~(TTCFs) of the operators on the environment part of \(H\Interaction\) completely determine the reduced dynamics of the system, i.e.~the reduced density matrix \(\rho\System(t)\).
This can be seen by either using the influence functional~\cite{Cirio:fermionic_influence_functional} or through the Keldysh formalism of Green's functions~\cite{Dorda:equivalence_fermionic_pseudomodes1,Chen2019}.
This fact has most useful consequences that we are going to illustrate, by means of some relevant examples, in the following subsections.
We will show, in particular, how two-time correlation functions provide the pivotal element which allows us to replace the original OQS environment(s) with transformed environments that are equivalent (i.e.~leading to the same open-system dynamics) but simpler to deal with.
As we will see next, such transformation essentially amounts to suitably absorbing the thermal factors, namely the average occupation, into the coupling between the system and  extended environments~\cite{Tamascelli2019:ttedopa}.

\subsection{Thermo-chemical spectral modulation} \label{sec:chemo-thermal}
Let us start by considering a simplified version of the model described by \cref{eq:fullHam,eq:hamiltonian_quadratic_env,eq:intHam}.
A fermionic quantum system $\System*$ interacts with a single environment $\Environment*$ consisting of a continuum of fermionic modes $\maybeadj{f_\omega}$, $\omega\in\Omega$ and initially set in a thermal state $\rho\Environment(0)$ at inverse temperature $\beta$ in the presence of a chemical potential $\mu$, i.e.
\begin{equation}
  \begin{aligned}
    H\Environment &= \int_\Omega \dd\omega\,(\omega - \mu) \adj{f_\omega} f_\omega\phantomadj,\\
    \rho\Environment(0) &= \frac{1}{\partitionf}\exp(-\beta H\Environment).
  \end{aligned} \label{eq:env_init_one}
\end{equation}
In this simplified setting, the system-bath interaction Hamiltonian can be written as
\begin{equation}
  \begin{split}
    H\Interaction&=
    \adj{A\System}\int_{\Omega}\dd\omega\,h(\omega)f_\omega
    -A\System\int_{\Omega}\dd\omega\,h(\omega)\adj{f_\omega}=\\&=
    \adj{A\System}B\Environment\phantomadj - A\System\phantomadj\adj{B\Environment},
  \end{split}
  \label{eq:sys_env_int}
\end{equation}
where $h(\omega)=\sqrt{J(\omega)}$, with $J\colon\Omega \to \R^+$ the spectral density, and $A\System$ is an arbitrary operator acting on the system.
The reduced dynamics of the system is therefore determined only by the following two-time correlation functions:
\begin{equation}
  \begin{aligned}
    c_0(t_1,t_2)&=
    \avg{ B\Environment(t_1)\adj{B\Environment(t_2)} }_{\rho\Environment(0)}=\\&=
    \Tr\Environment\bigl( B\Environment(t_1)\adj{B\Environment(t_2)}\rho\Environment(0)\bigr),\\
    c_1(t_1,t_2)&=
    \avg{ \adj{B\Environment(t_1)}B\Environment(t_2)}_{\rho\Environment(0)}=\\&=
    \Tr\Environment\bigl(\adj{B\Environment(t_1)}B\Environment(t_2)\rho\Environment(0)\bigr),
  \end{aligned}
\end{equation}
since the other two functions, which are $\avg{ B\Environment(t_1){B\Environment(t_2)} }_{ \rho\Environment(0)}$ and $ \avg{ \adj{B\Environment(t_1)}\adj{B\Environment(t_2)} }_{ \rho\Environment(0)} $, are identically equal to zero.
Here and in what follows $B\Environment(t)$ indicates the $B\Environment$ operator evolved under the free environment Hamiltonian $H\Environment$, i.e.~$B\Environment(t)\defeq e^{itH\Environment}B\Environment e^{-itH\Environment}$.
After very simple algebra, we obtain
\begin{equation}
  \begin{aligned}
    c_0(t_1,t_2) &= \int_{\Omega} \dd\omega\, J(\omega) \bigl(1-n_{\beta,\mu}(\omega)\bigr) e^{-i(\omega-\mu)(t_1-t_2)},\\
    c_1(t_1,t_2) &= \int_{\Omega} \dd\omega\, J(\omega) n_{\beta,\mu}(\omega)  e^{i(\omega-\mu)(t_1-t_2)}.
  \end{aligned}
  \label{eq:thermofield_cf}
\end{equation}
We notice that up until here the chemical potential has the only role of a frequency shift in the free Hamiltonian.
Without loss of generality, therefore, we can translate the frequencies so that the environment has \(\mu=0\); the support $\Omega$ of the spectral density becomes, consequently, $\Omega' = \Omega-\mu$.
Clearly enough, due to the stationarity of the initial state under the free evolution of the environment, the two functions depend on the time instants only through their difference \(t_1-t_2\).
In what follows we will thus define and use $ c_0(t)= c_0(t,0)$  and $ c_1(t)= c_1(t,0)$.

In close analogy with the T-TEDOPA strategy~\cite{Tamascelli2019:ttedopa}, we now introduce the modified spectral densities
\begin{equation}
  \begin{aligned}
    J_{\beta,\mu}^{(0)}(\omega)&\defeq
    (1-n_{\beta}(\omega))J(\omega+\mu),\\
    J_{\beta,\mu}^{(1)}(\omega)&\defeq
    n_{\beta}(\omega)J(\omega+\mu),
  \end{aligned}
  \label{eq:chemo_thermalizedSD}
\end{equation}
where \(n_\beta=n_{\beta,0}\), and \(J_{\beta,\mu}^{(j)}\colon \Omega' \to \R^+\), \(j\in\{0,1\}\).
We will refer to this transformation of the spectral density, accounting for the temperature and chemical potential terms, as the \emph{thermo-chemical spectral modulation}~(TCSM).
By rewriting \cref{eq:thermofield_cf} in terms of the thermo-chemically modulated spectral densities~\eqref{eq:chemo_thermalizedSD}
\begin{equation}
  \begin{aligned}
    c_0(t) &= \int_{\Omega'} \dd\omega\, J^{(0)}_{\beta,\mu}(\omega) e^{-it\omega},\\
    c_1(t) &= \int_{\Omega'} \dd\omega\, J^{(1)}_{\beta,\mu}(\omega) e^{it\omega},
  \end{aligned}
  \label{eq:tweigthTTCFs}
\end{equation}
it is possible to recognize that $c_0(t)$ coincides with the (only non-vanishing) TTCF
\begin{equation}
  \int_{\Omega'} \dd\omega\, J_{\beta,\mu}(\omega)\avg{ g_{0,\omega}\phantomadj(t) \adj{g_{0,\omega}} }_{\vacstate\vacstate*}
\end{equation}
of a fermionic environment of  modes $\maybeadj{g_{0,\omega}}$ initially set in the pure vacuum state $\vacstate \vacstate*$, with $g_{0,\omega}\vacstate = 0$, evolving under the free Hamiltonian $H_{\Environment*,0} = \int_{\Omega'} \dd \omega\,\omega\adj{g_{0,\omega}} g_{0,\omega}$ and interacting with the system via
\begin{equation}
  H_{\Interaction*,0} = \int_{\Omega'} \dd \omega\,\sqrt{J_{\beta,\mu}^{(0)}(\omega)}(\adj{A\System} g_{0,\omega}\phantomadj - A\System \phantomadj\adj{g_{0,\omega}}).
  \label{eq:intHamSingleZero}
\end{equation}
We can analogously identify $c_1(t)$ as
\begin{equation}
  c_1(t) = \int_{\Omega'} \dd \omega\, J_{\beta,\mu}(\omega) \avg{ \adj{g_{1,\omega}}(t) g_{1,\omega}\phantomadj }_{\filledstate \filledstate*}
\end{equation}
i.e.~the TTCF of a fermionic environment of modes $\maybeadj{g_{1,\omega}}$ initially set in the pure filled state $\filledstate \filledstate*$, with $\adj{g_{1,\omega}} \filledstate = 0$, evolving under the free Hamiltonian $H_{\Environment*,1} = \int_{\Omega'} \dd \omega\,\omega\adj{g_{1,\omega}} g_{1,\omega}\phantomadj$ and interacting with the system via
\begin{equation}
  H_{\Interaction*,1} = \int_{\Omega'} \dd \omega\,\sqrt{J_{\beta,\mu}^{(1)}(\omega)} (\adj{A\System} g_{1,\omega}\phantomadj - A\System\phantomadj \adj{g_{1,\omega}} ).
  \label{eq:intHamSingleOne}
\end{equation}

The two just introduced distinct environments, one starting in the pure vacuum state and the other in the pure filled state, and interacting with the system as
\begin{equation}
  \begin{split}
    H\Interaction'&=
    \begin{multlined}[t]
      \int_{\Omega'}\dd\omega\,\sqrt{J^{(0)}_{\beta,\mu}(\omega)}(\adj{A\System}g_{0,\omega}\phantomadj+\adj{g_{0,\omega}}A\System\phantomadj)\\
      +\int_{\Omega'}\dd\omega\,\sqrt{J^{(1)}_{\beta,\mu}(\omega)}(\adj{A\System}g_{1,\omega}\phantomadj+\adj{g_{1,\omega}}A\System\phantomadj)=
    \end{multlined}\\ &=
    \begin{multlined}[t]
      \adj{A\System}\int_{\Omega'}\dd\omega\,\bigl(
        \sqrt{J^{(0)}_{\beta,\mu}(\omega)}g_{0,\omega}+
        \sqrt{J^{(1)}_{\beta,\mu}(\omega)}g_{1,\omega}
      \bigr)\\
      -A\System \int_{\Omega'}\dd\omega\,\bigl(
        \sqrt{J^{(0)}_{\beta,\mu}(\omega)}\adj{g_{0,\omega}}+\sqrt{J^{(1)}_{\beta}(\omega)}\adj{g_{1,\omega}}
      \bigr)=
    \end{multlined}\\ &=
    \adj{A\System}B'\Environment-A\System\phantomadj\adj{{B'\Environment}}
  \end{split}
\end{equation}
determine the same reduced dynamics of the system $\System*$.
As a matter of fact, once defined the initial state \(\ket{\psi_0}\) such that \(g_{0,\omega}\ket{\psi_0}=0\) and \(\adj{g_{1,\omega}}\ket{\psi_0}=0\) for all \(\omega\in\Omega\), one can easily see that
\begin{equation}
  \begin{aligned}
    c_0'(t) &= \bra{\psi_0}{B'\Environment(t)}{{B'\Environment}}^\dagger\ket{\psi_0}=c_0(t),\\
    c_1'(t) &= \bra{\psi_0}{{B'\Environment(t)}}^\dagger{B'\Environment}\ket{\psi_0}=c_1(t).
  \end{aligned}
\end{equation}

This derivation of  auxiliary environments determining the same open-system dynamics leads to the same auxiliary environments obtained by the thermofield transformation~\cite{schwarz18} (see also \cref{sec:thermofield} for more details).
Differently from this latter approach, which is based on a purification of the thermal state and a suitable Bogoliubov transformation on an extended space of environmental modes, our approach pivots on the equivalence of auxiliary environments at the level of TTCFs.
Furthermore, by applying a particle-hole inversion \(f_{1,\omega}\phantomadj \to \adj{f_{1,\omega}}\) the variant of the fermionic thermofield transformation proposed in Ref.~\onlinecite{deVega15} is obtained.

The same derivation can be applied to interaction Hamiltonians of the form~\eqref{eq:intHam}, i.e.~where the open system is composed of multiple levels, each one coupling differently to the environments, if the hybridization functions for each level differ only by a constant, namely
\begin{equation}
    H\Interaction=
    \sum_{\lambda=1}^{m} \int_{\Omega} \dd \omega\,\kappa_\lambda \sqrt{J(\omega)} (\adj{A_{\lambda}}f_{\omega}\phantomadj+\adj{f_{\omega}}A_{\lambda}\phantomadj)
    \label{eq:sum_level_dependent_system_operators}
\end{equation}
(we use a single environment in this example for simplicity).
With the replacement $\tilde{A}\defeq \sum_{\lambda=1}^{m} \kappa_\lambda A_\lambda$ it is evident that, even in this case, one still ends up with the same (pure vacuum and pure filled) environments as in the case of a single system operator $A\System$; the interaction Hamiltonians~\eqref{eq:intHamSingleZero} and~\eqref{eq:intHamSingleOne} become
\begin{equation}
  \begin{split}
    H_{\Interaction*,j} &=
    \int_{\Omega'} \dd \omega\,\sqrt{J_{\beta,\mu}^{(j)}(\omega)} \sum_{\lambda=1}^{m} \kappa_\lambda (\adj{A_\lambda} g_{j,\omega}\phantomadj - A_\lambda\phantomadj \adj{g_{j,\omega}} )=\\ &=
    \int_{\Omega'} \dd\omega\,\sqrt{J_{\beta,\mu}^{(j)}(\omega)} (\adj{\tilde{A}} g_{j,\omega}\phantomadj - \tilde{A} \adj{g_{j,\omega}} ).
  \end{split}
  \label{eq:intHamSingleEq}
\end{equation}
with \(j\in\{0,1\}\).

\begin{figure*}
  \centering
  \begingroup
\pgfmathsetseed{2}
\newcommand*\circled[1]{%
  \tikz[baseline=(char.base)]{%
    \node[shape=circle,draw,inner sep=2pt] (char) {#1};
  }%
}
\newcommand{\omegaminL}{-0.5}
\newcommand{\omegamaxL}{1.5}
\newcommand{\temperatureL}{1}
\newcommand{\omegaminR}{-1}
\newcommand{\omegamaxR}{1}
\newcommand{\temperatureR}{0.25}
\newcommand{\shiftright}{2}
\newcommand{\shiftleft}{-5}
\newcommand{\scalefactor}{0.75}
\newcommand{\colpercentageL}{20}
\newcommand{\colpercentageR}{40}
\newcommand{\colpercentagecombined}{30}
\newcommand{\nptsleft}{4}
\newcommand{\nptsright}{5}
\newcommand{\ptsleft}{{-0.08, -0.52}, {0.83, 0.26}, {0.35, -0.52}, {-0.25, 0.21}}
\newcommand{\ptsright}{{-0.64, -0.27}, {0.79, -0.37}, {0.29, 0.58}, {0.23, 0.02}, {-0.83, 0.42}}
\newlength{\gridheight}%
\newlength{\gridwidth}%
\setlength{\gridheight}{0.18\textwidth}%
\setlength{\gridwidth}{0.23\textwidth}%
\newlength{\plotshiftx}
\newlength{\plotshifty}%
\setlength{\plotshiftx}{\dimexpr (0.5\gridwidth +5mm) / (-2) \relax}%
\setlength{\plotshifty}{-0.25\gridheight}%
\def\NE{8}
\def\residualN{4}
\tikzset{
  pic shift/.store in=\shiftcoord,
  pic shift={(0,0)},
  declare function = {
    JL(\x) = 2/pi * sqrt((\x - \omegaminL) * (\omegamaxL - \x));
    JR(\x) = 2/pi * sqrt((\x - \omegaminR) * (\omegamaxR - \x));
    nL(\x) = 0.5 * (1 - tanh(0.5 * (\x) / \temperatureL));
    nR(\x) = 0.5 * (1 - tanh(0.5 * (\x) / \temperatureR));
  },
  fill fraction/.style n args={3}{%
    path picture={
      \fill[#1] (path picture bounding box.south west) rectangle
      ($(path picture bounding box.south east)!#3!(path picture bounding box.north
      east)$);
      \fill[#2] ($(path picture bounding box.south west)!#3!(path picture bounding box.north
      west)$) rectangle (path picture bounding box.north east);
    }
  }
}%
\input{drawings_environment_addition_insets.tikz}%

\pgfdeclarelayer{arrow layer}%
\pgfdeclarelayer{ellipses layer}%
\pgfdeclarelayer{background}%
\pgfsetlayers{arrow layer,ellipses layer,background,main}%
\pgfmathsetseed{4} 
\begin{tikzpicture}
  \begin{scope}[on background layer]
    \matrix[
      matrix of nodes, nodes in empty cells,
      inner sep = 0pt,
      minimum width = \gridwidth,
      minimum height = \gridheight,
    ] (grid) 
    {
      &&&\\
      &&&\\
      &&&\\
      &&&\\
    };
  \end{scope}
  \begin{scope}[
      shift=($(grid-1-2)!0.5!(grid-1-3)$),
      local bounding box=original,
      name prefix=original-
    ]
    \node (sys) at (0,0) [system];
    \foreach \point [expand list, count=\n] in {\ptsleft} {
      \pgfmathsetmacro{\level}{random()}
      \node [circle, draw, fill fraction={gray}{white}{\level}] (el\n) at ($(sys)-(2,0)+(\point)$) {};
      \begin{pgfonlayer}{background}
        \draw (sys) -- (el\n);
      \end{pgfonlayer}
    }
    \foreach \point [expand list, count=\n] in {\ptsright} {
      \pgfmathsetmacro{\level}{random()}
      \node [circle, draw, fill fraction={gray}{white}{\level}] (er\n) at ($(sys)+(2,0)+(\point)$) {};
      \begin{pgfonlayer}{background}
        \draw (sys) -- (er\n);
      \end{pgfonlayer}
    }
    \begin{pgfonlayer}{ellipses layer}
      \node [fit=(el1) (el2) (el3) (el4), ellipse, fill=violet!\colpercentageL!white] (leftenv) {};
      \node [fit=(er1) (er2) (er3) (er4) (er5), ellipse, fill=violet!\colpercentageR!white] (rightenv) {};
    \end{pgfonlayer}
    \pic [pic shift={(\omegaminR+\plotshiftx,\plotshifty)}] at (grid-1-1) {origLsdf};
    \pic [pic shift={(\omegaminR+\plotshiftx,\plotshifty)}] at (grid-1-4) {origRsdf};
  \end{scope}
  \begin{scope}[
      local bounding box=split,
      name prefix=split-,
      shift=($(grid-2-2)!0.5!(grid-3-3)$)
    ]
    \node (sys) at (0,0) [system];
    \foreach \point [expand list, count=\n] in {\ptsleft} {
      \node [circle, draw, fill=white] (elh\n) at ($(sys)+(-2,+1.5)+(\point)$) {};
      \node [circle, draw, fill=gray]  (elc\n) at ($(sys)+(-2,-1.5)+(\point)$) {};
      \begin{pgfonlayer}{background}
        \draw (sys) -- (elh\n);
        \draw (sys) -- (elc\n);
      \end{pgfonlayer}
    }
    \foreach \point [expand list, count=\n] in {\ptsright} {
      \node [circle, draw, fill=white] (erh\n) at ($(sys)+(+2,+1.5)+(\point)$) {};
      \node [circle, draw, fill=gray]  (erc\n) at ($(sys)+(+2,-1.5)+(\point)$) {};
      \begin{pgfonlayer}{background}
        \draw (sys) -- (erh\n);
        \draw (sys) -- (erc\n);
      \end{pgfonlayer}
    }
    \begin{pgfonlayer}{ellipses layer}
      \node [fit=(elh1) (elh2) (elh3) (elh4), ellipse, fill=strongred!\colpercentageL!white] (lefthenv) {};
      \node [fit=(elc1) (elc2) (elc3) (elc4), ellipse, fill=strongblue!\colpercentageL!white] (leftcenv) {};
      \node [fit=(erh1) (erh2) (erh3) (erh4) (erh5), ellipse, fill=strongred!\colpercentageR!white] (righthenv) {};
      \node [fit=(erc1) (erc2) (erc3) (erc4) (erc5), ellipse, fill=strongblue!\colpercentageR!white] (rightcenv) {};
    \end{pgfonlayer}
    \pic [pic shift={(\omegaminR+\plotshiftx,\plotshifty)}] at (grid-2-1) {splithotLsdf};
    \pic [pic shift={(\omegaminR+\plotshiftx,\plotshifty)}] at (grid-3-1) {splitcoldLsdf};
    \pic [pic shift={(\omegaminR+\plotshiftx,\plotshifty)}] at (grid-2-4) {splithotRsdf};
    \pic [pic shift={(\omegaminR+\plotshiftx,\plotshifty)}] at (grid-3-4) {splitcoldRsdf};
  \end{scope}
  \begin{scope}[
      local bounding box=combined,
      name prefix=combined-,
      shift=($(grid-4-2)!0.5!(grid-4-3)$)
    ]
    \node (sys) at (0,0) [system];
    \foreach \point [expand list, count=\n] in {\ptsleft} {
      \node [circle, draw, fill=white] (elh\n) at ($(sys)+(-2,0)+(\point)$) {};
      \node [circle, draw, fill=gray]  (elc\n) at ($(sys)+(+2,0)+(\point)$) {};
      \begin{pgfonlayer}{background}
        \draw (sys) -- (elh\n);
        \draw (sys) -- (elc\n);
      \end{pgfonlayer}
    }
    \foreach \point [expand list, count=\n] in {\ptsright} {
      \node [circle, draw, fill=white] (erh\n) at ($(sys)+(-2,0)+(\point)$) {};
      \node [circle, draw, fill=gray]  (erc\n) at ($(sys)+(+2,0)+(\point)$) {};
      \begin{pgfonlayer}{background}
        \draw (sys) -- (erh\n);
        \draw (sys) -- (erc\n);
      \end{pgfonlayer}
    }
    \begin{pgfonlayer}{ellipses layer}
      \node [fit=(elh1) (elh2) (elh3) (elh4) (erh1) (erh2) (erh3) (erh4) (erh5), ellipse, fill=strongred!\colpercentagecombined!white] (henv) {};
      \node [fit=(elc1) (elc2) (elc3) (elc4) (erc1) (erc2) (erc3) (erc4) (erc5), ellipse, fill=strongblue!\colpercentagecombined!white] (cenv) {};
    \end{pgfonlayer}

    \pic [pic shift={(\omegaminR+\plotshiftx,\plotshifty)}] at (grid-4-1) {combinedhotsdf};
    \pic [pic shift={(\omegaminR+\plotshiftx,\plotshifty)}] at (grid-4-4) {combinedcoldsdf};
  \end{scope}
  \begin{scope}[
      shift={($(combined-sys)+(0,-3)$)},
      local bounding box = tedopabox,
      name prefix=tedopa-
    ]
    \node (e0) at (0,0) [system, alias=eh0, alias=ec0];
    \pgfmathsetmacro{\NElong}{int(\NE+\residualN)}
    \foreach \n [remember=\n as \m (initially 0)] in {1,...,\NElong} {
      \node [two level system, draw=black, fill=spinblue] (ec\n) at ($(e0)+2/3*(\n,0)$) {};
      \node [two level system, draw=black, fill=spinred]  (eh\n) at ($(e0)-2/3*(\n,0)$) {};
      \draw (ec\m) -- (ec\n);
      \draw (eh\m) -- (eh\n);
    }
    \draw [dashed, dash pattern={on 1mm off 1mm}] (ec\NElong) -- +(+0.75,0);
    \draw [dashed, dash pattern={on 1mm off 1mm}] (eh\NElong) -- +(-0.75,0);
  \end{scope}
  \begin{pgfonlayer}{arrow layer}
    \draw [line width=10*\the\thicklwidth, shorten <=5mm, shorten >=4mm, snakearrowcolour, line cap=round, -{Straight Barb[round]}] (original-sys) .. controls +(-65:6cm) and ($(tedopa-e0)+(110:6cm)$) .. (tedopa-e0)%
    node [stronggreen, pos=0.03, left=2mm] {\circled{1}}%
    node [stronggreen, pos=0.36, right=5mm] {\circled{2}}%
    node [stronggreen, pos=0.72, right=1mm] {\circled{3}}%
    node [stronggreen, pos=1, above right=1mm] {\circled{4}};
  \end{pgfonlayer}
\end{tikzpicture}
\endgroup
  \caption{%
    (1) The initial system is coupled to two leads, each one starting from a thermal state with different temperature and chemical potential (here we show the two spectral density functions already shifted).
    (2) Each lead is split into two new environments, one starting from the empty state, the other from the filled state, with thermo-chemically modulated spectral densities.
    (3) The two empty environments are merged, summing their spectral densities; the same is done with the filled environments.
    (4) The resulting environments are chain-mapped into discrete sets of fermionic modes with nearest-neighbor interaction.%
  }
  \label{fig:baths_chain_mapping_thermofield}
\end{figure*}

\subsection{Environment additivity} \label{sec:additivity}
The previous results can be exploited as a starting point to derive auxiliary environments for more complex scenarios.
We extend the system introduced in the previous subsection by considering the interaction of the open system $\System*$ with two separated leads, modeled as fermionic baths labeled by $\Left*$ and $\Right*$.
We will consider the general case, with the two leads possibly having different temperature, chemical potential and system-bath interaction strength profiles, as defined by the respective spectral densities $J_{\Left*(\Right*)}(\omega)$.
The system $\System*$, therefore, connects the leads and acts as a bridge to transport energy and/or particles.

The TCSM introduced in the previous subsection can be used to replace each of the finite-temperature environments with two pure state auxiliary environments.
The resulting configuration thus comprises four separated pure state environments $\Environment*_{\alpha,k}$, with \(\alpha\in\{\Left*,\Right*\}\) and \(k\in\{0,1\}\), each one interacting with the system with a Hamiltonian
\begin{equation}
  H_{\Interaction*,(\alpha,k)}=
  \int_{\Omega_\alpha'}\dd \omega\,\sqrt{J_{\beta_\alpha,\mu_\alpha}^{(k)}(\omega)}  (\adj{A\System} g_{(\alpha,k),\omega}\phantomadj - A\System\phantomadj \adj{g_{(\alpha,k),\omega}}),
\end{equation}
with $\Omega_\alpha' = \Omega_\alpha - \mu_\alpha$ and the auxiliary environments $\Environment*_{\alpha,0}$  and  $\Environment*_{\alpha,1}$ in the vacuum and filled state respectively.
This configuration can be simplified; as a matter of fact  a direct inspection of the system-bath interaction Hamiltonian
\begin{multline}
  H\Interaction=
  \adj{A\System}B\Left\phantomadj
  -A\System\phantomadj\adj{B\Left} + \adj{A\System}B\Right\phantomadj
  -A\System\phantomadj\adj{B\Right}
  =\adj{A\System} C -A\System\phantomadj\adj{C}
 \label{eq:sys_env_int_LR}
\end{multline}
where $\maybeadj{B_\alpha} = \int_{\Omega_\alpha}\dd \omega\, \sqrt{J_\alpha(\omega)}\maybeadj{f_{\alpha,\omega}}$ and $\maybeadj{C} = \maybeadj{B\Left}+ \maybeadj{B\Right}$, reveals that, as in the case of a single bath, there are only two non-vanishing TTCFs and that each of them receives contributions from both the left and the right lead.
For example, if we indicate by $\rho\Environment(0) = \rho\Left(0)\rho\Right(0)$ the (factorized) initial state of the leads, each one starting from a thermal state at inverse temperature $\beta_{\Left*(\Right*)}$, we have
\begin{equation}
  \begin{split}
    \avg{ C(t) \adj{C} }_{\rho\Environment(0)} &=  \avg{ B\Left\phantomadj(t) \adj{B\Left} }_{\rho\Left(0)} +\avg{ B\Right\phantomadj(t) \adj{B\Right} }_{\rho\Right(0)}=\\ &=
    \int_{\Omega_\text{ext}} \dd \omega\,e^{-i \omega t} \sum_{\alpha\in\{\Left*,\Right*\}} J_{\beta_\alpha,\mu_\alpha}^{(0)} (\omega) \indicatorf{\Omega_\alpha'}(\omega)
  \end{split} \label{eq:c0_LR}
\end{equation}
where $\Omega_\text{ext}\defeq\Omega\Left' \cup \Omega\Right'$, $J_{\beta,\alpha}^{(0)}$ is defined as in \cref{eq:chemo_thermalizedSD} and $\indicatorf{\Omega'}$ is the indicator function of the set $\Omega'=\Omega-\mu$.
In the last line of \cref{eq:c0_LR} it is easy to recognize the TTCF corresponding to a fermionic bath of modes $\maybeadj{g_{0,\omega}}$ in the frequency range $\Omega_\text{ext}$, initially set in the vacuum state, undergoing a free evolution determined by $H_{\Environment*,0}=\int_{\Omega_\text{ext}}\dd\omega\,\omega \adj{g_{0,\omega}} g_{0,\omega}\phantomadj$ and interacting with the open system $\System*$ through the Hamiltonian
\begin{equation}
  H_{\Interaction*,0}'= \int_{\Omega_\text{ext}}\dd \omega\,\sqrt{J_\text{ext}^{(0)}(\omega)} (\adj{A\System} g_{0,\omega}\phantomadj - A\System\phantomadj \adj{g_{0,\omega}}),
\end{equation}
with
\begin{equation}
  J_\text{ext}^{(0)}(\omega)\defeq \sum_{\alpha\in\{\Left*,\Right*\}} J_{\beta_\alpha,\mu_\alpha}^{(0)} (\omega) \indicatorf{\Omega_\alpha'}(\omega).
  \label{eq:extTwoVacuum}
\end{equation}
In a similar way we can see that the TTCF $\avg{ \adj{C(t)} {C} }_{\rho\Environment(0)}$ is determined by a collection of modes $\maybeadj{g_{1,\omega}}$ starting in the filled state and interacting with the system via
\begin{equation}
  H_{\Interaction*,1}'= \int_{\Omega_\text{ext}}\dd \omega\,\sqrt{J_\text{ext}^{(1)}(\omega)} (\adj{A\System} g_{1,\omega}\phantomadj - A\System\phantomadj \adj{g_{1,\omega}} ),
\end{equation}
where
\begin{equation}
  J_\text{ext}^{(1)}(\omega)\defeq\sum_{\alpha\in\{\Left*,\Right*\}} J_{\beta_\alpha,\mu_\alpha}^{(1)} (\omega) \indicatorf{\Omega_\alpha'}(\omega).
  \label{eq:extTwoFull}
\end{equation}

We have therefore shown that the two leads, each starting in a thermal state in the presence of (possibly different) chemical potentials, can be replaced by two equivalent pure state environments, one starting from the vacuum state and the other starting from the filled state, and each accounting for thermo-chemically weighted contributions from the two leads by means of modified spectral densities $J_\text{ext}^{(k)}$ for $k\in\{0,1\}$.
An extension to multiple environments or to the case of a level-dependent system-bath coupling follows immediately (see \cref{eq:intHamSingleEq}).
We finally observe that the environment additivity property can be applied to purified environments as obtained, for example, by means of the thermofield approach~\cite{schwarz18}.
It is still under investigation whether there exists a way to resolve the dynamics pertaining to a single environment among the merged ones.
So far we have not found any, and the answer to the question seems to be no, which would mean that after the merging step the information about the individual environments is lost.

\subsection{Modified resonant level model and spin bath}
Let us consider a special instance, also known as the \emph{modified resonant level model}~\cite{Nuesseler2020:ftedopa}, of the system studied in \cref{sec:chemo-thermal}.
The interaction between the fermionic open system $\System*$ and a single environment defined as in \cref{eq:env_init_one}, is now given by
\begin{equation} \label{eq:hintself}
  H\Interaction = i A\System \int_\Omega \dd \omega\, \sqrt{J(\omega)} B_\omega,
\end{equation}
with $A\System$ and $B_{\omega}$ \emph{self-adjoint} operators acting, respectively, on the system and on the bath side. 
We observe that, because of the CARs, the system and bath operators anticommute, so we need an additional \(i\) factor in the definition of \(H\Interaction\) to make it Hermitian.
We moreover assume that the operators $B_{\omega}$ are linear in the creation and annihilation operators $\adj{f_\omega}$ and $f_\omega$.
In what follows we will set $B_{\omega} =  X_\omega = f_\omega\phantomadj + \adj{f_\omega}$, but our results will be valid for any self-adjoint linear combination $aX_{\omega}+bP_\omega$, i.e.~of the (Majorana) operators $X_\omega$ and $P_\omega = i(\adj{f_\omega}-f_\omega\phantomadj)$.

In this setting, the reduced dynamics of the system is determined by a single TTCF, namely
\begin{equation*}
  \begin{aligned}
    &C(t) = \int_\Omega \dd\omega\, J(\omega) \avg{X_\omega(t) X_\omega (0)}_{\rho_{\beta,\mu}(\omega)}=\\
    &=\int_\Omega \dd\omega\, J(\omega) \bigl[e^{-i(\omega-\mu) t} \bigl(1-n_{\beta,\mu}(\omega)\bigr) \!+\!e^{i(\omega-\mu) t} n_{\beta,\mu}(\omega)\bigr]
  \end{aligned}
\end{equation*}
with $\rho_{\beta,\mu}$ and $n_{\beta,\mu}$ defined as in \cref{eq:rhobetamu,eq:avocc} respectively.
The following identity, moreover, holds:
\begin{equation} \label{eq:ttcf1}
  \begin{split}
    C(t)&=
    \int_{\Omega} \dd\omega\, J(\omega)\avg{ X_\omega(t) X_\omega (0)}_{\rho_{\beta,\mu}(\omega)}=\\
    &=\begin{multlined}[t]
      \int_{\Omega'} \dd\omega\, J(\omega+\mu )\bigl(1-n_{\beta}(\omega)\bigr) e^{-i\omega t}\\
      +\int_{-\Omega'} \dd\omega'\, J(-\omega'+\mu)n_{\beta}(-\omega') e^{-i\omega' t}
    \end{multlined}
  \end{split}
\end{equation}
where we have applied the shift $\omega \to \omega-\mu$, so that $\Omega' = \Omega-\mu$, as in \cref{sec:chemo-thermal}, $n_\beta = n_{\beta,0}$, and we used the change of variable $\omega' = -\omega$ in the second integral.
Since
\begin{equation}
  n_{\beta}(-\omega)=
  1-n_{\beta}(\omega)=
  \frac12\Bigl(1+\tanh\frac{\beta\omega}{2}\Bigr),
\end{equation}
we can rewrite the whole integral as the transform of a single function \(J\Ext_{\beta,\mu}\) defined as
\begin{multline}
  J\Ext_{\beta,\mu}(\omega)\defeq
  \frac12\Bigl(1+\tanh\frac{\beta\omega}{2}\Bigr)\bigl[
    \indicatorf{-\Omega'}(\omega)J(-\omega+\mu)\\ 
    +\indicatorf{\Omega'}(\omega)J(\omega+\mu)
  \bigr],
  \label{eq:extSD}
\end{multline}
so that
\begin{equation}
  C(t)=
  \int_{-\infty}^{+\infty}\dd\omega\,J\Ext_{\beta,\mu}(\omega) e^{-i\omega t}.
  \label{eq:ttcf2}
\end{equation}
In \cref{eq:ttcf2} it is easy to recognize the TTCF of an environment consisting of modes $\maybeadj{g}_\omega$, $\omega\in\R$ each starting from the vacuum state $\ket{0}_\omega$, freely evolving under $H_{\Environment*,\omega} = \omega \adj{g_\omega} g_\omega\phantomadj$ and interacting with the system as \(H_{\Interaction*,\omega} =\sqrt{J\Ext_{\beta,\mu}(\omega)} A\System X_\omega\).

In the presence of multiple baths, e.g.~left and right fermionic leads, with different temperatures, chemical potentials and spectral densities, we can apply the same construction to each bath independently and exploit, similarly to what we did in \cref{sec:additivity}, the additivity of the TTCF to derive a single equivalent environment, always starting from the pure vacuum state, with an extended spectral density equal to the sum of the spectral densities of the transformed separated environments.
For example, if we consider an environment comprising two fermionic leads, as in \cref{eq:hamiltonian_quadratic_env}, interacting with the system via the interaction term
\begin{equation}
  H\Interaction=
  i \sum_{\lambda=1}^{m}\sum_{\alpha\in\{\Left*,\Right*\}}  \int_{\Omega_\alpha}\dd \omega \,h_{\lambda,\alpha}(\omega) A_\lambda X_{\alpha,\omega}, \label{eq:selfadjint}
\end{equation}
with a self-adjoint $A_\lambda$, $X_{\alpha,\omega} = f_{\alpha,\omega}\phantomadj + \adj{f_{\alpha,\omega}}$ and $h_{\lambda,\alpha}(\omega) =  \kappa_\lambda \sqrt{J_{\alpha}(\omega)}$, $J_\alpha\colon\Omega_\alpha \to \R^+$, our construction will define an equivalent environment comprising modes $\maybeadj{g}_\omega$, $\omega \in \R$ each starting from the vacuum state $\ket{0}_\omega$, freely evolving under $H_{\Environment*,\omega} = \omega \adj{g_\omega} g_\omega\phantomadj$ and interacting with the system as
\begin{equation}
  H_{\Interaction*,\omega}^{\text{ext}} =   \sum_{\lambda=1}^{m} \kappa_\lambda A_\lambda \sqrt{J^{\text{ext}}(\omega)}  X_\omega,
\end{equation}
where
\begin{align}
  J^{\text{ext}}(\omega) = J^{\text{ext}}_{\beta\Left}(\omega)+ J^{\text{ext}}_{\beta\Right}(\omega), \label{eq:extSD2}
\end{align}
with $J^{\text{ext}}_{\beta_\alpha}(\omega)$, $\alpha = \{\Left*,\Right*\}$, defined as in \cref{eq:extSD}.
This result extends the one obtained in Ref.~\onlinecite{Nuesseler2020:ftedopa}, where only the zero chemical potential case was considered.

The results of this section allow to extend the T-TEDOPA derivation introduced in Ref.~\onlinecite{Tamascelli2019:ttedopa} for single bosonic environments to the case of multiple environment at different temperature and (negative) chemical potentials.
As a matter of fact, once the thermal factors have been shifted from the initial states of the bosonic modes to the spectral density, all the environments start from the same (vacuum) state, and their action on the system is the same as a single environment over an extended support $\Omega_\text{ext}$ and system-bath interaction strengths modeled by a spectral density of the form~\eqref{eq:extSD2}.

Another relevant application of thermo-chemical spectral density modulation is the study of open quantum systems interacting with (multiple) spin baths in the thermodynamic limit.
In order to understand why it is the case, it is expedient to consider a single spin environment consisting of $N$ spin-$1/2$ particles.
Each spin evolves under the local Hamiltonian
\begin{equation}
  H^\text{spin}_j = {\epsilon_j} \sigma_j^+ \sigma_j^-,
  \label{eq:freeSpinH}
\end{equation}
where $\sigma_j^k$, $k\in\{x,y,z\}$ are the Pauli operators acting on the $j$-th spin, $\sigma_{\pm}^j = (\sigma_x^j \pm i \sigma_y^j)/2$ and we assume, without loss of generality, that $\epsilon_j \in \Omega = [0,\omega_\text{max}]$.
Each spin starts from the thermal state $\rho_j^\beta = \exp(-\beta H_j)/\partitionf_j$, where $\partitionf_j$ is the partition function; the system bath interaction has the form
\begin{equation}
  H\Interaction^{\text{spin}} = \sum_{j=1}^ N h_j A\System \sigma_j^x =
  \sum_{j=1}^ N h_j A\System (\sigma_j^++\sigma_j^-).
  \label{eq:intSpinH}
\end{equation}
\Cref{eq:freeSpinH,eq:intSpinH} become in the thermodynamic limit \(N\to+\infty\)
\begin{align}
  H\Environment^{\text{spin}} &= \int_{\Omega} \dd \omega\,\omega \sigma_+^\omega \sigma_-^\omega, \\
  H\Interaction^{\text{spin}} &= A\System \int_{\Omega} \dd \omega\,h(\omega) \sigma_{x}^{\omega}.
\end{align}
As shown in Refs.~\onlinecite{Makri1999:linear_response_influence_functional, Cao2018:spinorbit_iridates}, in this limit the spin bath behaves as a Gaussian  fermionic bath, so that, if the initial state is a Gaussian state, the reduced dynamics of the system is completely determined by the TTCF\@.
It is thus clear that all the results presented in this and in the previous subsections apply to spin baths in the thermodynamic limit as well.

\begin{figure}
  \centering
  \begingroup
\newcommand{\temperature}{0.3}
\newcommand{\chempot}{0.6}
\newcommand{\omegamax}{2}

\begin{tikzpicture}
  \pgfmathsetmacro{\rlim}{\omegamax-\chempot}
  \datavisualization [
    school book axes,
    visualize as line/.list = {lo, mi, hi, semicircle},
    lo = {style={line cap=round, strongred, line width=\plotlwidth}},
    mi = {style={line cap=round, strongred, line width=\plotlwidth}},
    hi = {style={line cap=round, strongred, line width=\plotlwidth}},
    semicircle = {style={dashed, violet}},
    x axis = {
      label  = $\omega$,
      length = \plotwidth,
      ticks  = {major={at={
            -\rlim    as $-\omega\Max+\mu$,
            -\chempot as $-\mu$,
            0         as $\vphantom{-}0$,
            \chempot  as $\vphantom{-}\mu$,
            \rlim     as $\omega\Max-\mu$,
            \omegamax as $\vphantom{-}\omega\Max$,
      }}},
    },
    y axis = {
      length    = \plotheight,
      ticks     = {major={at={
            0,
      }}},
    }
  ]
  data [set=lo, format=function] {
    var x : interval [-\rlim:-\chempot] samples 50;
    func y = 1/pi * (1 + tanh(0.5 * \value{x} / \temperature)) * sqrt((\chempot - \value{x}) * (\omegamax - \chempot + \value{x}));
  }
  data [set=mi, format=function] {
    var x : interval [-\chempot:\chempot-0.02] samples 50;
    func y = 1/pi * (1 + tanh(0.5 * \value{x} / \temperature)) * (
      sqrt((\chempot - \value{x}) * (\omegamax - \chempot + \value{x})) +
      sqrt((\chempot + \value{x}) * (\omegamax - \chempot - \value{x}))
    );
  }
  info {
    \node (Jext) at (visualization cs: x=-\chempot, y=0.35)
    [strongred] {$J\Ext_{\beta,\mu}(\omega)$};
  }
  data [set=mi, format=function] {
    var x : interval [\chempot-0.02:\chempot] samples 201;
    func y = 1/pi * (1 + tanh(0.5 * \value{x} / \temperature)) * (
      sqrt((\chempot - \value{x}) * (\omegamax - \chempot + \value{x})) +
      sqrt((\chempot + \value{x}) * (\omegamax - \chempot - \value{x}))
    );
  }
  data [set=hi, format=function] {
    var x : interval [\chempot:\rlim-0.01] samples 50;
    func y = 1/pi * (1 + tanh(0.5 * \value{x} / \temperature)) * sqrt((\chempot + \value{x}) * (\omegamax - \chempot - \value{x}));
  }
  data [set=hi, format=function] {
    var x : interval [\rlim-0.01:\rlim] samples 201;
    func y = 1/pi * (1 + tanh(0.5 * \value{x} / \temperature)) * sqrt((\chempot + \value{x}) * (\omegamax - \chempot - \value{x}));
  }
  data [set=semicircle, format=function] {
    var x : interval [0:\omegamax-0.02] samples 50;
    func y = 2/pi * sqrt((\omegamax - \value{x}) * \value{x});
  }
  info {
    \node (J) at (visualization cs: x=1.75, y=0.6)
    [violet] {$J(\omega)$};
  }
  data [set=semicircle, format=function] {
    var x : interval [\omegamax-0.02:\omegamax] samples 101;
    func y = 2/pi * sqrt((\omegamax - \value{x}) * \value{x});
  };
\end{tikzpicture}
\endgroup
  \caption{Example of an extended spectral density resulting from \cref{eq:extSD}, starting from a semicircle spectral density on \((0,\omega\Max)\).}
  \label{fig:semicircle_tftedopa}
\end{figure}

\section{TCSM-TEDOPA} \label{sec:tcsm-tedopa}
In the previous section we showed how the thermo-chemical spectral density modulation leads to the definition of equivalent environments starting from pure vacuum/filled states and to the recombination of different baths.
In this section we will show the impact of this approach on non perturbative simulations of open quantum system dynamics by means of the TEDOPA algorithm.
We refer the reader to Refs.~\onlinecite{chin10, woods14, Nuesseler2020:ftedopa, Tamascelli2019:ttedopa} for a full account on TEDOPA for bosonic and fermionic environments.

To introduce the TEDOPA method it is convenient to start from the system defined in \cref{sec:chemo-thermal}: a fermionic system $\System*$ interacting with a fermionic bath $\Environment*$ according to \cref{eq:sys_env_int}, with the free Hamiltonian and initial state of the bath defined as in \cref{eq:env_init_one}.
A spectral density $J\colon\Omega \to\R^+$ which is absolutely continuous with respect to the Lebesgue measure defines a measure
\begin{equation}
  \dd\lambda(\omega)=J(\omega)\,\dd\omega;
\end{equation}
if the spectral density moreover belongs to the Szegő class then there exists a unique family of polynomials $\{p_n(\omega)\}_{n\in\N}$ which is orthogonal with respect to the measure $\dd \lambda(\omega)$~\cite{chin10}.
We refer the reader to \cref{sec:szego} for a precise definition of the Szegő class.

By exploiting the properties of such orthogonal polynomials (in particular the three-term recurrence relation they satisfy) it is possible to introduce a unitary transformation $U_n(\omega)$ on the environmental fermionic modes $\adj{f_\omega}$ resulting in a countable set of new fermionic modes $\maybeadj{c_n}$, $n\in\N$, satisfying the CAR $\{c_n,\adj{c_m}\} = \delta_{n,m}$.
The Hamiltonian \eqref{eq:fullHam} can be therefore unitarily transformed as follows: 
\begin{equation}
  \begin{aligned}
    H_{\System*\Environment*}\Chain&=
    H\System+H\Interaction\Chain+H\Environment\Chain\, ,\\
    H\Interaction\Chain&=
    \eta (\adj{A\System}c_0\phantomadj + \adj{c_0}A\System\phantomadj)\, , \\ 
    H\Environment\Chain&=
    \sum_{n=0}^{+\infty}\omega_n\adj{c_n}c_n\phantomadj+ \sum_{n=0}^{+\infty}\kappa_n(\adj{c_{n+1}}c_n\phantomadj+\adj{c_n}c_{n+1}\phantomadj) \nonumber,
    \label{eq:chainMapHam}
  \end{aligned}
\end{equation}
where the coefficients $\eta$, $\omega_n$ and $\kappa_n$ depend on the spectral density $J(\omega)$.
Note that the interaction strength between the system and the first chain mode \(\eta = \sqrt{\int_\Omega\dd\omega\,J(\omega)}\) is proportional to the overall system-bath interaction strength, and does not depend on any particular features of the spectral density.
Moreover different (bilinear) interaction Hamiltonians would result in the same chain Hamiltonians $H\Environment\Chain$ and different system-chain interactions $H\Interaction\Chain$.

The determination of the chain initial state
\begin{equation}
  \rho\Environment\Chain(0)=
  \frac{\exp(-\beta H\Environment\Chain)}{\partitionf}
\end{equation}
is generally a non-trivial problem.
Even at zero temperature the TEDOPA mapping brings one major disadvantage: it builds new fermionic bath modes as a linear combination of all original bath modes, thereby mixing occupied and empty bath modes.
Hence, the Fermi sea is not a trivial product state in the chain geometry and is typically highly entangled.
Zero- and finite-temperature initial states of the chain are typically  (approximately) determined  by suitable projections on the ground or thermal state, respectively.
If, for example, the system-chain state were represented as an MPS, a computationally demanding DMRG search of the ground or thermal state would be required~\cite{wolf14}.
The results presented in the previous section, or the thermofield transformation, allow us to circumvent this issue.
The mapping procedure described above can be applied to the auxiliary environments as well, as long as the thermally weighted spectral densities $J_{\beta}^{(0)}(\omega)$ and $J_{\beta}^{(1)}(\omega)$ defined in \cref{eq:chemo_thermalizedSD} are in the Szegő class.
As shown in \cref{sec:szego}, this is indeed the case as long as the original spectral density $J(\omega)$ is Szegő.
The combination of the fermionic spectral density modulation and the TEDOPA chain mapping of fermionic environments results in what we call the Thermo-Chemical Spectral Density Modulation TEDOPA method (TCSM-TEDOPA).
For example, a system with an exchange-type interaction such as the one described by the Hamiltonian~\eqref{eq:sys_env_int} is mapped into a configuration where the open system interacts with two different fermionic chains, namely
\begin{equation}
  \begin{aligned}
    H\Chain &= H\System+H_{\Environment*,0}\Chain + H_{\Environment*,1}\Chain + H_{\Interaction*,0}\Chain + H_{\Interaction*,1}\Chain,\\
    H_{\Environment*,j}\Chain &= \sum_{n=0}^{+\infty}\omega_{j,n}\phantomadj\adj{c_{j,n} }c_{j,n} \phantomadj\!+\! \sum_{n=0}^{+\infty}\kappa_{j,n}\phantomadj(\adj{c_{j,n+1} }c_{j,n}\phantomadj+\adj{c_{j,n}}c_{j,n+1}\phantomadj),\\
    H_{\Interaction*,j}\Chain &=
  \eta_j (\adj{A\System} c_{j,0}\phantomadj+\adj{c_{j,0}} A\System\phantomadj),
  \end{aligned}
  \label{eq:cmapExc}
\end{equation}
where the coefficients $\eta_j$, $\omega_{j,n}$ and $\kappa_{j,n}$ depend on the thermalized spectral densities $J_{\beta}^{(j)}$ for $j=0,1$, and the modes $c_{0,n}$ and $c_{1,n}$ start, respectively, from the vacuum and filled state.

When the system interacts with two fermionic environments, as in the model discussed in \cref{sec:additivity}, one could proceed in a similar way and map each lead into two chains.
The resulting configuration would then have the system interacting with four chains.
While this topology can be represented by means of tensor networks, it would lead to a highly non-local entanglement structure, known to be severely detrimental to the efficiency of DMRG and time-evolution algorithms.
The results obtained in \cref{sec:additivity}, however, suggest a different approach.
As a matter of fact, if the spectral densities $J_{L}(\omega)$ and $J_{R}(\omega)$ are Szegő-class and $\Omega_L\cap\Omega_R\neq \emptyset$, i.e.~if the supports of the considered spectral densities, already suitably shifted by the corresponding chemical potentials, do overlap, then the extended spectral densities $J_\text{ext}^{(0)}(\omega)$ and $J_\text{ext}^{(1)}(\omega)$ defined in \cref{eq:extTwoFull,eq:extTwoVacuum} are Szegő as well.
This in turn means that only two chains, one  corresponding to the chain mapping of $J_\text{ext}^{(0)}(\omega)$ and starting from the vacuum state and one corresponding to the chain mapping of $J_\text{ext}^{(1)}(\omega)$ and starting from the filled state, are sufficient to determine the same reduced dynamics of the system.
We observe that the condition $\Omega_L\cap\Omega_R\neq \emptyset$ is indeed a very mild one, since in conduction schemes the leads do typically include the system transition frequency.
From the discussion of \cref{sec:additivity} it is clear that the same procedure can be extended, under the same conditions, to more than two environments and to the case of level-dependent  system-bath interaction strength, {by first summing the level-specific system-side operators within \(H\Interaction\) into a collective one as in \cref{eq:sum_level_dependent_system_operators}.}
As a matter of example, let us consider the general model introduced in \cref{sec:model}.
The application of the TCSM-TEDOPA procedure will lead to a Hamiltonian of the same form as \cref{eq:cmapExc} with
\begin{equation}
  H_{\Interaction*,j}\Chain =
  \eta_j(\adj{\tilde{A}\System} c_{j,0}\phantomadj+\adj{c_{j,0}} \tilde{A}\System\phantomadj),
\end{equation}
where
\begin{equation}
  \tilde{A}\System=\sum_{\lambda=1}^{m}\kappa_\lambda A_\lambda,\quad
  \eta_{j}\defeq\sqrt{\int_{\Omega_\text{ext}} \dd \omega\,J_\text{ext}^{(j)}(\omega)}.
\end{equation}
We lastly remark that, if the interaction Hamiltonian is of the form~\eqref{eq:selfadjint}, it is possible to use a \emph{single} chain, with chain coefficients determined by the spectral density $J_{\beta}\Ext(\omega)$ and system-bath and all the modes starting from the vacuum state.

\section{Fermionic Markovian closure}
\label{sec:fermionic_markovian_closure}
\newcommand{\minfreq}{\omega_\textnormal{m}}
\newcommand{\maxfreq}{\omega_\textnormal{M}}
\newcommand{\homfrequency}{\Omega}
\newcommand{\homcoupling}{K}
After the chain mapping procedure has been applied, we are left with the system interacting with semi-infinite chain(s) of fermionic modes.
It is clear that in actual simulations the chain(s) needs to be truncated after a certain number of sites.
In this section we will restrict our attention to the single chain case; we will show towards the end of the section how the results extend to the multiple chain case.

The truncation point, i.e.~the number $N$ of chain sites that are kept, must be suitably chosen so as not to induce finite-size effects on the system and on the chain dynamics.
It is therefore clear that such a choice depends on both the speed at which the perturbations induced by the interaction of the first chain site with the system propagate along the chain and on the simulation time $t_\text{max}$, so that $N = N(t_\text{max})$.
While the exact dependence on the simulation time is hard to define, in general for sufficiently large $t_\text{max}$ Lieb-Robinson bound techniques suggest that $N(t_\text{max})$ scales linearly in $t_\text{max}$.
We will better motivate this claim in the following subsection.
What is important to stress here is that longer simulation times require longer chains so that the determination of long-time dynamics can soon become prohibitively expensive.

The Markovian Closure (MC) mechanism, recently proposed in Ref.~\onlinecite{Nuesseler2022:markovian_closure} for bosonic environments, provides a solution to this problem.
In the following subsections we will introduce the basic ideas behind the MC and formulate an equivalent mechanism for fermionic environment.
The key ingredient in the MC is the equivalence theorem~\cite{Tamascelli2018:equivalence_theorem} which allows establishing when a unitarily evolving environment induces the same reduced dynamics on an open quantum system \(\System*\) as an auxiliary environment undergoing a non unitary Lindblad evolution.

\subsection{Asymptotic coefficients} \label{sec:asymptotic_coeff}
If a spectral density $J\colon [\minfreq,\maxfreq]\to\R^+$ belongs to the Szegő class, then the chain coefficients have an interesting convergence property~\cite{chin10,woods14}: the sequences of site energies and interaction coefficients \(\{\omega_n\}_{n=0}^{+\infty}\) and \(\{\kappa_n\}_{n=1}^{+\infty}\) converge, for $n\to+\infty$, to limiting values that depend only on the support of $J$:
\begin{align}
  \homfrequency&\defeq
  \lim_{n\to+\infty}\omega_n=
  \frac{\maxfreq+\minfreq}{2},\\
  \homcoupling&\defeq
  \lim_{n\to+\infty}\kappa_n=
  \frac{\maxfreq-\minfreq}{4}.
\end{align}
In analogy to what has been recently proposed in Ref.~\onlinecite{Nuesseler2022:markovian_closure} for the case of bosonic environments, once the coefficients have converged towards their asymptotic values up to a desired tolerance, say $\epsilon>0$, i.e.~$\abs{\omega_n - \homfrequency} < \epsilon$ and $\abs{\kappa_n - \homcoupling} < \epsilon$ for all $n>N\Environment = N\Environment(\epsilon)$, the chain part comprising the modes $\maybeadj{c_n}$ for $n>N\Environment$ can be approximated by a uniform chain whose coefficients  $\omega_n$ and $\kappa_n$ are set equal to $\homfrequency$ and $\homcoupling$ respectively.
The resulting chain Hamiltonian reads
\begin{multline} \label{eq:approxHam}
  \widetilde{H}\Environment\Chain=
  \sum_{n=0}^{N\Environment}
  \omega_n\adj{c_n}c_n\phantomadj
  +\sum_{n=0}^{N\Environment}
  \kappa_n(\adj{c_{n+1}}c_n\phantomadj+\adj{c_n}c_{n+1}\phantomadj)\\
  +\sum_{n=N\Environment+1}^{+\infty}
  \!\homfrequency\adj{c_n}c_n\phantomadj
  +\sum_{n=N\Environment+1}^{+\infty}
  \!\homcoupling(\adj{c_{n+1}}c_n\phantomadj+\adj{c_n}c_{n+1}\phantomadj).
\end{multline}
It helps the physical intuition to think of the uniform chain as a ``runway'': particles/holes entering this region are propagated without the possibility of being scattered back by inhomogeneities in the couplings or in the frequencies~\cite{tama20Entro}.
The propagation speed of a particle/hole over this uniform region is thus constant in time.
This justifies the $O(t_\text{max})$ dependence the truncation point $N$ used at the beginning of the section.

Such flat part of the chain can be traced back (by inverting the chain mapping) to a continuous fermionic environment, that we call \emph{residual environment}, characterized by
\begin{equation}
  H_0\Residual=
  \int_{\minfreq}^{\maxfreq} \dd\omega\,\omega\adj{f_\omega}f_\omega\phantomadj,
  \label{eq:residual_environment_free}
\end{equation}
with \(\minfreq\defeq\homfrequency-2\homcoupling\) and \(\maxfreq\defeq\homfrequency+2\homcoupling\),
interacting with the \(N\Environment\)-th mode (the last one of the inhomogeneous part) through
\begin{equation}
  H\Interaction\Residual=
  \int_{\minfreq}^{\maxfreq}\dd\omega\,\sqrt{J_\infty(\omega)}(\adj{c_{N\Environment}}f_\omega\phantomadj+\adj{f_\omega}c_{N\Environment}\phantomadj),
  \label{eq:residual_environment_interaction}
\end{equation}
where \(J_\infty\) is the spectral density
\begin{equation}
  \begin{aligned}
    J_{\infty}&\colon[\homfrequency-2\homcoupling,\homfrequency+2\homcoupling]\to\R^+\\
    \omega&\mapsto
    \frac{1}{2\pi}\sqrt{(2\homcoupling-\homfrequency+\omega)(2\homcoupling+\homfrequency-\omega)}.
  \end{aligned}
  \label{eq:semicircle_spectral_density}
\end{equation}
The convergence toward a flat residual spectrum can be seen as an embedding of the original
system into an enlarged set of degrees of freedom that evolve in a Markovian way~\cite{martinazzo11}.
In fact, in Ref.~\onlinecite{Nuesseler2022:markovian_closure} it was proven that a bosonic residual environment having spectral density $J_\infty$ can be replaced with an auxiliary system, the Markovian Closure, that consists of a small number of damped, interacting bosonic modes {undergoing a Lindbladian evolution}.
This mechanism allows for the replacement of a semi-infinite homogeneous chain of harmonic modes with a finite auxiliary environment acting as an absorber for the excitations traveling along the homogeneous part of the  chain.
The MC is moreover \emph{universal}, in the sense that it can be applied to all chain mappings of bosonic environments with spectral densities in the Szegő class.
In what follows we will provide an analogous construction for fermionic environments.
More in detail, we will show how the semi-infinite homogeneous chain of fermionic modes governed by the Hamiltonian 
\begin{equation}
  H\sys{hom}=
  \sum_{n=N\Environment+1}^{+\infty}\!\homfrequency
  \adj{c_n}c_n\phantomadj+
  \sum_{n=N\Environment+1}^{+\infty}\!\homcoupling
  (\adj{c_{n+1}}c_n\phantomadj+\adj{c_n}c_{n+1}\phantomadj)
\end{equation}
from \cref{eq:approxHam} can be replaced by a fermionic Markovian closure having the same structure as the one devised for bosonic systems.
The resulting closure, moreover, has the same universality character of its bosonic counterpart: it can be applied to all fermionic environments with Szegő spectral density.

\subsection{Derivation of the fermionic Markovian Closure}
\label{ssec:bath-pseudomodes_equivalence_theorem}
Ref.~\onlinecite{Tamascelli2018:equivalence_theorem} shows how the reduced dynamics of an open system in contact with a bosonic bath can, under certain conditions, be obtained by replacing the bath with a suitably parameterized (finite) set of pseudomodes, i.e.~damped harmonic oscillators.
An equivalent result for fermionic systems is provided by Ref.~\onlinecite{Chen2019}, in which the original bath is first mapped, as we did in \cref{sec:fermionic_spectral_density_thermalization}, to an auxiliary system comprising \emph{two} different baths, one of which is initially completely filled (\(\mu=+\infty\)) and the other completely empty (\(\mu=-\infty\)), representing the original environmental modes below and above the chemical potential.
The reduced dynamics of the open system, with this new auxiliary configuration, is then equated to one where the environment is a set of damped fermionic oscillators, with appropriate parameters.
Indeed, the equivalence of the reduced dynamics generated by the different environments rests on the equivalence of the correlation functions of the interaction Hamiltonian.

We start, as we did in \cref{sec:chemo-thermal}, by defining
\begin{equation}
  A\System \defeq c_{N\Environment},
  \quad
  B\Environment \defeq \int_{\minfreq}^{\maxfreq} \dd\omega\,\sqrt{J_\infty(\omega)}f_\omega
\end{equation}
so that we can write the interaction Hamiltonian~\eqref{eq:residual_environment_interaction} as
\begin{equation} \label{eq:hint_res_comp}
  H\Interaction\Residual=
  \adj{A\System}B\Environment\phantomadj+\adj{B\Environment}A\System\phantomadj.
\end{equation}
We must then compute the correlation functions
\begin{equation}
  \begin{aligned}
    c_0(t) &= \avg{B\Environment(t)\adj{B\Environment(0)}},\\
    c_1(t) &= \avg{\adj{B\Environment(t)}B\Environment(0)}
  \end{aligned}
  \label{eq:residual_environment_correlation_functions}
\end{equation}
for the initially empty environment, which starts from the vacuum state \(\vacstate\); we have \(B\Environment\vacstate=0\), so \(c_1(t)\) is identically zero and we are left only with
\begin{equation}
  c_0(t)= \vacstate* B\Environment(t)\adj{B\Environment(0)} \vacstate =
  \int_{\minfreq}^{\maxfreq}\dd\omega\, e^{-i\omega t} J_\infty(\omega).
\end{equation}
When the environment starts from a completely filled state \(\filledstate\) instead, as the other chain does, \(\adj{B\Environment}\filledstate=0\) so we only have
\begin{equation}
  c(t)=
  \filledstate* \adj{B\Environment(t)}B\Environment(0) \filledstate=
  \int_{\minfreq}^{\maxfreq}\dd\omega\, e^{i\omega t}J_\infty(\omega).
\end{equation}
By computing the Fourier transform of \(J_\infty\) we finally get
\begin{equation}
  \begin{aligned}
    c_0(t)&=
    \homcoupling^2e^{-i\homfrequency t}C\semicircle(2Kt),\\
    c_1(t)&=
    \homcoupling^2e^{i\homfrequency t}C\semicircle(2Kt),
  \end{aligned}
\end{equation}
where
\begin{equation}
  C\semicircle(x)=
  \int_{-1}^{1}\dd y\,e^{-ixy}\frac{2}{\pi}\sqrt{1-y^2}=
  J_0(x)+J_2(x)
\end{equation}
is the transform of the unit-radius semicircle spectral density
\begin{equation}
  j\semicircle(y) = \frac{2}{\pi}\sqrt{1-y^2}
  \label{eq:unitsc}
\end{equation}
and \(J_n(x)\) denote Bessel functions of the first kind.

\newcommand{\PMfreqmat}{\Lambda}%
\newcommand{\PMdissmat}{\Gamma}%
\newcommand{\PMfreqmatemp}{\PMfreqmat^{(0)}}%
\newcommand{\PMdissmatemp}{\PMdissmat^{(0)}}%
\newcommand{\PMfreqmatfil}{\PMfreqmat^{(1)}}%
\newcommand{\PMdissmatfil}{\PMdissmat^{(1)}}%
Pseudomodes are a new set of fermionic modes, which we describe by a different set of fermionic annihilation and creation operators \(\maybeadj{a_j}\).
Let us start from the residual environment associated to a chain in the vacuum state: this system acts as a perfect absorber of excitations, so we set
its initial state as the vacuum \(\vacstate\vacstate*\) and that its free dynamics is generated by the Lindblad operator \(\lindblad\Reservoir^{(0)}(\rho)=-i\bcomm{H\Reservoir^{(0)},\rho}+\dissipator\Reservoir^{(0)}(\rho)\) with
\begin{equation}
  H\Reservoir^{(0)}=
  \sum_{i,j=1}^{N\Closure}\PMfreqmatemp_{ij}\adj{a_i}a_j\phantomadj
  \label{eq:hamiltoniano_libero_pseudomodi}
\end{equation}
and
\begin{equation}
  \dissipator\Reservoir^{(0)}(\rho)=
  \sum_{i,j=1}^{N\Closure}\PMdissmatemp_{ij}\Bigl(a_j\phantomadj\rho \adj{a_i}-\frac12\fcomm{\rho,\adj{a_i}a_j\phantomadj}\Bigr);
  \label{eq:dissipatore_pseudomodi}
\end{equation}
note that the vacuum is stationary with respect to the evolution operator, i.e.~\(\lindblad\Reservoir^{(0)}(\vacstate\vacstate*)=0\).
We define a new operator \(B\Reservoir\), which plays the same role as \(B\Environment\) in the exchange-interaction Hamiltonian~\eqref{eq:hint_res_comp},
\begin{equation}
  B\Reservoir\defeq
  \sum_{k=1}^{N\Closure}\zeta_ka_k,
  \label{eq:interazione_pseudomodi}
\end{equation}
for some \(\zeta_k\in\C\).
This operator generates new correlation functions,
\begin{equation}
  \begin{aligned}
    c'_0(t) &= \avg{B\Reservoir(t)\adj{B\Reservoir(0)}},\\
    c'_1(t) &= \avg{\adj{B\Reservoir(t)}B\Reservoir(0)},
  \end{aligned} \label{eq:aux_TTCF}
\end{equation}
that must match, respectively, \(c_0(t)\) and \(c_1(t)\) from \cref{eq:residual_environment_correlation_functions} if we want the reduced dynamics to be equal.
We refer the reader to \cref{sec:FMC_details} for full detail on the determination of the equivalent auxiliary environment;
here we limit ourselves to report the fundamental steps.

A particular solution to our problem exists for the unit-radius semicircle spectral density $j\semicircle$ defined in \cref{eq:unitsc}.
In detail, let
\begin{equation}
  M=
  \begin{pmatrix}
    \alpha_1 & \beta_1  & 0        & 0        & \cdots & 0 \\
    \beta_1  & \alpha_2 & \beta_2  & 0        & \cdots & 0 \\
    0        & \beta_2  & \alpha_3 & \beta_3  & \cdots & 0 \\
    0        & 0        & \beta_3  & \alpha_4 & \cdots & 0 \\
    \vdots   & \vdots   & \vdots   & \vdots   & \ddots & \vdots \\
    0        & 0        & 0        & 0        & \cdots & \alpha_{N\Closure}
  \end{pmatrix} \label{eq:mmatrix}
\end{equation}
be a \(N\Closure\times N\Closure\) complex matrix, and \(\vec{w}\in\C^{N\Closure}\) be a solution to
\begin{equation} \label{eq:problem}
  \innerp{ w , \exp(tM)w } =
  C\semicircle(t).
\end{equation}
It then follows that by setting, in \cref{eq:hamiltoniano_libero_pseudomodi,eq:dissipatore_pseudomodi},
\begin{equation}
  \PMfreqmatemp=
  \begin{pmatrix}
    \omega_1  & g_1      & 0        & \cdots & 0 \\
    g_1      & \omega_2 & g_2      & \cdots & 0 \\
    0        & g_2      & \omega_3 & \cdots & 0 \\
    \vdots   & \vdots   & \vdots   & \ddots & \vdots \\
    0        & 0        & 0        & \cdots & \omega_{N\Closure}
  \end{pmatrix}\, \label{eq:lambdazero}
\end{equation}
and
\begin{equation}
  \PMdissmatemp_{i,j} =  \gamma_j \delta_{i,j}, \label{eq:gammazero}
\end{equation}
where
\begin{equation}
  \begin{aligned}
    \omega_j &= \homfrequency-2\homcoupling\Im\alpha_j,\\
    g_j      &= -2\homcoupling\Im\beta_j,\\
    \gamma_j &=  -4\homcoupling\Re\alpha_j ,
  \end{aligned}
  \label{eq:mc_parameters}
\end{equation}
and by choosing, in \cref{eq:interazione_pseudomodi},
\begin{equation}
  \zeta_j = \homcoupling w_j,
\end{equation}
the equality
\begin{align}
  c_0(t)&=
  \homcoupling^2e^{-i\homfrequency t}C\semicircle(2Kt)
  = \avg{B\Reservoir(t)\adj{B\Reservoir(0)}} = c'_0(t)
\end{align}
holds for all \(t \geq 0\).
The auxiliary environment defined by \cref{eq:interazione_pseudomodi,eq:lambdazero,eq:gammazero} is a collection of $N\Closure$ pseudomodes, with nearest-neighbor interaction, undergoing local dissipation and each starting from the vacuum state and interacting with the truncation site $N\Environment$ of the chain; this provides a \emph{fermionic Markovian closure} (FMC) for a TEDOPA chain of fermionic modes starting from the vacuum state $\vacstate$.
If the filled state $\filledstate$ is the initial state of the chain, we only have to use \(\conj{\alpha_j}\) and \(\conj{\beta_j}\) instead of \(\alpha_j\) and \(\beta_j\) respectively, and set the initial state of the closure sites to the filled state as well (see \cref{sec:FMC_details}).
In \cref{fig:chain_mapping_with_mc} we provide a graphical representation of the TCSM-TEDOPA transformation and of the finite-size system resulting from the use of the FMC construct.
\begin{figure}
  \centering
  \begingroup
\def\NE{4}%
\def\NC{6}%
\def\NElong{6}%
\def\modestretch{0.55}%

\begin{tikzpicture}
  \clip (-0.5\columnwidth,3.2mm) rectangle +(\columnwidth,-135.30682pt);
  \begin{scope}[local bounding box=tedopa, name prefix=tedopa-]
    \node (e0) at (0,0) [system, alias=eh0, alias=ec0, alias=snakept];
    \foreach \n [remember=\n as \m (initially 0)] in {1,...,\NElong} {
      \node [two level system, draw=black, fill=spinblue] (ec\n) at ($(e0)+\modestretch*(\n,0)$) {};
      \node [two level system, draw=black, fill=spinred]  (eh\n) at ($(e0)-\modestretch*(\n,0)$) {};
      \draw (ec\m) -- (ec\n);
      \draw (eh\m) -- (eh\n);
    }
    \draw [dashed, dash pattern={on 1mm off 1mm}] (ec\NElong) -- +(+1.5*\modestretch,0);
    \draw [dashed, dash pattern={on 1mm off 1mm}] (eh\NElong) -- +(-1.5*\modestretch,0);
  \end{scope}
  \begin{scope}[local bounding box=pseudomodebox, yshift=-2.5cm, name prefix=pseudomodes-]
    \node (e0) at (0,0) [system, alias=eh0, alias=ec0];
    \foreach \n [remember=\n as \m (initially 0)] in {1,...,\NE} {
      \node [two level system, draw=black, fill=spinblue] (ec\n) at ($(ec0)+\modestretch*(\n,0)$) {};
      \node [two level system, draw=black, fill=spinred] (eh\n) at ($(eh0)-\modestretch*(\n,0)$) {};
      \draw (eh\m) -- (eh\n);
      \draw (ec\m) -- (ec\n);
    }
    \foreach \n [evaluate=\n as \angle using -\pseudomodeanglespan+2*\pseudomodeanglespan*(\n-1)/(\NC-1)] in {1,...,\NC} {
      \node [two level system, draw=black, fill=lightspinred]  (sh\n) at ($(eh\NE)-(\angle:\pseudomodeleg)$) {};
      \node [two level system, draw=black, fill=lightspinblue] (sc\n) at ($(ec\NE)+(\angle:\pseudomodeleg)$) {};
      \draw (eh\NE) -- (sh\n);
      \draw (ec\NE) -- (sc\n);
      \draw [decorate, decoration={snake, amplitude=\snakeamp, segment length=\snakeper}] (sh\n.west) -- +(-\snakelen,0) coordinate (sh\n-end);
      \draw [decorate, decoration={snake, amplitude=\snakeamp, segment length=\snakeper}] (sc\n.east) -- +(\snakelen,0) coordinate (sc\n-end);
    }
    \foreach \n [remember=\n as \m (initially 1)] in {2,...,\NC} {
      \draw (sh\m) -- (sh\n);
      \draw (sc\m) -- (sc\n);
    }
    \coordinate (lbd) at ([yshift=-2ex]sh\NC.south);
    \pgfmathsetmacro{\midpm}{int(\NC/2)}
    \draw ([yshift=\bracketdrop]{eh1.south east |- lbd}) -- (eh1.south east |- lbd) to node [midway, below] {\scriptsize initial state: completely empty} (sh\midpm.west |- lbd) -- ([yshift=\bracketdrop]{sh\midpm.west |- lbd});
    \draw ([yshift=\bracketdrop]{ec1.south west |- lbd}) -- (ec1.south west |- lbd) to node [midway, below] {\scriptsize initial state: completely filled} (sc\midpm.east |- lbd) -- ([yshift=\bracketdrop]{sc\midpm.east |- lbd});
    \coordinate (cred) at (ec2.north);
    \coordinate (cblue) at (eh2.north);
  \end{scope}
  \begin{scope}[on background layer]
    \draw [line width=5*\the\thicklwidth, shorten <=2ex, shorten >=2ex, snakearrowcolour, line cap=round, -{Straight Barb[round, length=2ex]}] (tedopa-e0) .. controls ([shift={(1,-1.5)}]tedopa-e0) and ([shift={(-1.5,2)}]pseudomodes-e0) .. (pseudomodes-e0);
  \end{scope}
\end{tikzpicture}
\endgroup%
  \caption{%
    The chain-mapped system before and after truncating the TEDOPA chains and replacing the residual environments, on both the initially empty and initially filled side, with appropriate sets of interacting pseudomodes.%
  }
  \label{fig:chain_mapping_with_mc}
\end{figure}

We can see in \cref{eq:mc_parameters} that the parameters needed to define the Markovian closure modes depend on the original environments only through \(\homfrequency\) and \(\homcoupling\), which are easily obtainable from the lower and upper bounds of the domain of the original spectral density function.
In order to implement the fermionic Markovian closure it is therefore only necessary to apply an affine rescaling to the \(\alpha_j\) and \(\beta_j\) coefficients, a feature shared with the bosonic Markovian closure~\cite{Nuesseler2022:markovian_closure}.
Determining the \(\alpha_j\) and \(\beta_j\) coefficients requires the solution of a non-trivial inversion problem~\cite{mascherpa20} and, in general, only approximate solutions (e.g.~by means of approximated Prony methods~\cite{beylkin2005}) can be found.
As shown in \cref{sec:FMC_details}, however, we do not need to solve the problem anew, but we can use the same coefficients determined in Ref.~\onlinecite{Nuesseler2022:markovian_closure} for the bosonic setting.
We refer the reader to the Supplemental Material of the same work for an analysis of the accuracy of the approximation, which holds here as well.
For the sake of self-containedness the parameters for closures consisting of $N\Closure=6,8,10$ modes are provided anyway in \cref{sec:FMC_details}.

Finally, we remark that the semicircular density of states is not the only way we can reshape an environment.
Other fitting schemes have been proposed, e.g.~in Ref.~\onlinecite{Dorda2017:nonmarkovian_impurity_problems_Lindblad} or Ref.~\onlinecite{Cirio:fermionic_influence_functional}, targeting specific spectral densities.
While a tailored fitting procedure might produce, by directly replacing the original environment with some pseudomodes and skipping the TEDOPA construction altogether, equivalent environments with a smaller number of modes, the construction of the pseudomodes with the Markovian closure is a more generic approach which does not rely on particular features of the initial spectral densities.

\subsection{Error sources}
We observe, moreover, that different interaction patterns, e.g.~with fully connected pseudomodes or next-nearest neighbor interactions, and alternative methods for the determination of the closure coefficients, as the one recently proposed in Ref.~\onlinecite{feist21,feist21nano}, could be used to either reduce the number of modes in the closure or further reduce the approximation error.

Another source of errors is the choice of the convergence point $N\Environment$, i.e.~the chain site connected to the FMC\@.
As a matter of fact, as remarked in \cref{sec:asymptotic_coeff}, the chain coefficient $\omega_n$ and $\kappa_n$ converge to the values $\homfrequency$ and $\homcoupling$ only asymptotically, so that the actual residual spectral density is only approximated by a semicircle spectral density $J_\infty$.
For any assigned spectral density, however, it is always possible to make a suitable choice  of $N\Environment$ and to estimate the corresponding error~\cite{mascherpa17}.
As we will see in the next sections, moreover, in the case of spectral densities that are relevant in the fermionic setting, the approximate convergence is reached within a very small number ($N\Environment < 10)$ of chain sites.

\section{Numerical tests}
\label{sec:numerical_tests}
With the application of the Markovian closure construction, the system resulting from the TCSM mapping is no longer one-dimensional; the small number of pseudomodes usually needed to represent the residual environments, however, makes it still amenable to such a representation.
As a matter of fact, it suffices to ``flatten'' the pseudomodes into a linear configuration at the cost of creating non-nearest-neighbor interactions between the sites (see \cref{fig:tensor_network_scheme}).
We point out that our procedure cannot be, in general, used if the open system is composed of multiple levels each of which interacts with substantially different environments, i.e.~with different spectral density functions that are not multiples of one another, or through different operators on the system side in the interaction Hamiltonian.
In the presence of \(m\) environments, the number of auxiliary environments would scale as \(2m\), and they cannot be efficiently merged in just one or two chains since the resulting local dimension would grow exponentially with \(m\).

Moreover, we encode the state into a matrix-product state, instead of a matrix-product operator, by first vectorizing the density matrix, i.e.~by taking its representative vector in a basis which is orthonormal with respect to the Hilbert-Schmidt inner product, and then we compute its time evolution by means of the time-dependent variational principle~(TDVP) algorithm~\cite{haege16}.
\begin{figure}
  \centering
  \begingroup
\def\NC{4}
\def\NE{3}
\def\scalef{0.2}

\begin{tikzpicture}[
    scale           = \scalef,
    overhang/.style = {shorten <=\overhang, shorten >=\overhang},
  ]
  \node (el0) at (0,0) [system square, alias=er0];
  \foreach \n [remember=\n as \m (initially 0)] in {1,...,\NE} {
    \node [square, draw=black, fill=spinblue] (el\n) at ($(el0)-(2.5*\n,0)$) {};
    \draw [strongblue] (el\m) -- (el\n);
    \node [square, draw=black, fill=spinred]  (er\n) at ($(er0)+(2.5*\n,0)$) {};
    \draw [strongred] (er\m) -- (er\n);
  }
  \foreach \n [evaluate=\n as \m using \NE+\n] in {1,...,\NC} {
    \node [square, draw=black, fill=lightspinblue] (sl\n) at ($(el\NE)-(2.5*\n,0)$) {};
    \node [square, draw=black, fill=lightspinred]  (sr\n) at ($(er\NE)+(2.5*\n,0)$) {};
    \pgfmathsetmacro{\angleout}{-60*(1-4^(1-\n))}
    \pgfmathsetmacro{\anglein} {180-\angleout}
    \draw [overhang, strongblue, out=\anglein, in=\angleout] (el\NE) to (sl\n);
    \draw [overhang, strongred, out=\angleout, in=\anglein]  (er\NE) to (sr\n);
  }
  \foreach \n [remember=\n as \m (initially 1)] in {2,...,\NC} {
    \draw [strongblue] (sl\m) -- (sl\n);
    \draw [strongred]  (sr\m) -- (sr\n);
  }
  \coordinate (lbd) at ([yshift=-1cm]er0.south);
  \foreach \n [remember=\n as \m (initially 0)] in {1,...,\NE} {
    \node [square, draw=black, fill=spinblue] at (el\n) {};
    \node [square, draw=black, fill=spinred] at (er\n) {};
  }
  \foreach \n [evaluate=\n as \angle using \NE+\n] in {1,...,\NC} {
    \node [square, draw=black, fill=lightspinblue] at (sl\n) {};
    \node [square, draw=black, fill=lightspinred] at (sr\n) {};
  }
\end{tikzpicture}
\endgroup
  \begingroup
\def\NC{4}
\def\NE{3}
\def\scalef{0.2}

\begin{tikzpicture}[
    scale           = \scalef,
    overhang/.style = {shorten <=\overhang, shorten >=\overhang},
  ]
  \node (el0) at (0,0) [system square, alias=er0];
  \foreach \n [remember=\n as \m (initially 0)] in {1,...,\NE} {
    \node [square, draw=black, fill=spinblue] (el\n) at ($(el0)+({2.5+5*(\n-1)},0)$) {};
    \node [square, draw=black, fill=spinred]  (er\n) at ($(er0)+(5*\n,0)$) {};
  }
  \pgfmathsetmacro{\angleout}{45}
  \pgfmathsetmacro{\anglein} {180-\angleout}
  \draw [overhang, strongblue, out=\angleout, in=\anglein]  (el0.east |- el1.north) to (el1);
  \draw [overhang, strongred, out=-\angleout, in=-\anglein] (er0.east |- er1.south) to (er1);
  \foreach \n [remember=\n as \m (initially 1)] in {2,...,\NE} {
    \path [spath/save global=cl\n, overhang, out=45, in=135]  (el\m) to (el\n);
    \path [spath/save global=cr\n, overhang, out=-45, in=-135] (er\m) to (er\n);
  }
  \foreach \n in {1,...,\NC} {
    \node [square, draw=black, fill=lightspinblue] (sl\n) at ($(el\NE)+(5*\n,0)$) {};
    \pgfmathsetmacro{\anglein} {-45*(1-10^(-\n))}
    \pgfmathsetmacro{\angleout}{180-\anglein}
    \path [spath/save global=pmol\n, overhang, out=\anglein, in=\angleout] (el\NE) to (sl\n);
    \node [square, draw=black, fill=lightspinred] (sr\n) at ($(er\NE)+(5*\n,0)$) {};
    \pgfmathsetmacro{\angleout}{-\anglein}
    \pgfmathsetmacro{\anglein} {180-\angleout}
    \path [spath/save global=pmor\n, overhang, out=\angleout, in=\anglein] (er\NE) to (sr\n);
  }
  \foreach \n [remember=\n as \m (initially 1)] in {2,...,\NC} {
    \pgfmathsetmacro{\angleout}{45}
    \pgfmathsetmacro{\anglein} {180-\angleout}
    \path [spath/save global=pmil\n, overhang, out=\angleout, in=\anglein]   (sl\m) to (sl\n);
    \path [spath/save global=pmir\n, overhang, out=-\angleout, in=-\anglein] (sr\m) to (sr\n);
  }
  \tikzset{
    spath/.cd,
    split at intersections with  = {cr3}{pmol1},
    split at intersections with  = {cr3}{pmol2},
    split at intersections with  = {cr3}{pmol3},
    split at intersections with  = {cr3}{pmol4},
    insert gaps after components = {cr3}{\crossinggap},
    split at intersections with  = {pmil2}{pmor1},
    insert gaps after components = {pmil2}{\crossinggap},
    split at intersections with  = {pmil3}{pmor2},
    insert gaps after components = {pmil3}{\crossinggap},
    split at intersections with  = {pmil4}{pmor3},
    insert gaps after components = {pmil4}{\crossinggap},
    split at intersections with  = {pmir2}{pmol2},
    insert gaps after components = {pmir2}{\crossinggap},
    split at intersections with  = {pmir3}{pmol3},
    insert gaps after components = {pmir3}{\crossinggap},
    split at intersections with  = {pmir4}{pmol4},
    insert gaps after components = {pmir4}{\crossinggap},
  }
  \foreach \n in {2,...,\NE} {
    \draw [spath/use=cl\n, strongblue];
    \draw [spath/use=cr\n, strongred];
  }
  \foreach \n in {1,...,\NC} {
    \draw [spath/use=pmol\n, strongblue];
    \draw [spath/use=pmor\n, strongred];
  }
  \foreach \n in {2,...,\NC} {
    \draw [spath/use=pmil\n, strongblue];
    \draw [spath/use=pmir\n, strongred];
  }
  \node at (0,0) [system square];
  \foreach \n in {1,...,\NE} {
    \node [square, draw=black, fill=spinblue] at (el\n) {};
    \node [square, draw=black, fill=spinred]  at (er\n) {};
  }
  \foreach \n in {1,...,\NC} {
    \node [square, draw=black, fill=lightspinblue] at (sl\n) {};
    \node [square, draw=black, fill=lightspinred]  at (sr\n) {};
  }
\end{tikzpicture}
\endgroup
  \caption{%
    \textit{Top}: tensor network layout representing the application of a Markovian closure on both chains of an environment to the chains resulting from TCSM-TEDOPA transformation.
    \textit{Bottom}: interleaved configuration in the tensor network. In both frames, a line between two squares denotes an interaction.%
  }
  \label{fig:tensor_network_scheme}
\end{figure}
Following Ref.~\onlinecite{Kohn2021:interleaved_mapping} we also change how the sites are enumerated in order to interleave the two environments, so that the open system is put at one edge of the linear chain (\cref{fig:tensor_network_scheme} at the bottom).
The two chains in the original scheme (\cref{fig:tensor_network_scheme} at the top) would, in fact, become strongly correlated over time, as a consequence of the evolution towards a steady state; therefore, the original ``naive'' tensor network develops long-range correlations between the two different pseudomode sets.
The interleaved configuration instead brings the two environments ``closer'' to each other, allowing us to use a lower overall bond dimension in the matrix-product states; this turns the original nearest-neighbor interactions into next-nearest-neighbor ones, which is an acceptable trade-off.

Throughout this section we will use the SIAM~\cite{anderson61} as an example model.
Generally this model describes electrons, so we have two fermionic levels \(\sigma\in\{\spinup,\spindown\}\) (representing the spin degrees of freedom) making up the system Hamiltonian
\begin{equation}
  H\System = \epsilon_{\spinup}\phantomadj\adj{d_{\spinup}}d_{\spinup}\phantomadj + \epsilon_{\spindown}\phantomadj\adj{d_{\spindown}}d_{\spindown}\phantomadj + U\adj{d_{\spinup}}d_{\spinup}\phantomadj\adj{d_{\spindown}}d_{\spindown}\phantomadj.\\
  \label{eq:siam_system_hamiltonian}
\end{equation}
In the continuum limit the environment---also composed of electrons---can be described by the following Hamiltonian operators:
\begin{equation}
  \begin{aligned}
    H\Environment &= \sum_{\sigma\in\{\spinup,\spindown\}}\int_\Omega\dd\omega\, \omega\adj{f_{\sigma,\omega}}f_{\sigma,\omega}\phantomadj,\\
    H\Interaction &= \sum_{\sigma\in\{\spinup,\spindown\}}\int_\Omega\dd\omega\, \sqrt{J(\omega)}(\adj{f_{\sigma,\omega}}d_\sigma\phantomadj+\adj{d_\sigma}f_{\sigma,\omega}\phantomadj)
  \end{aligned}
  \label{eq:siam_environment_hamiltonian}
\end{equation}
for some spectral density function \(J\), starting from the thermal state at a given temperature and chemical potential.
We will use for the most part the non-interacting version, i.e.~with \(U=0\), to test the fermionic MC method.
In this version different spins do not interact anymore (through the system), so we can consider a single spin component only: instead of electrons we will speak of \emph{spinless fermions}, i.e.~each mode will represent just a two level (empty/occupied) degree of freedom.

\subsection{Correlation functions from simulations}
First of all we check that the we can faithfully reproduce the correlation functions of the environment from the FMC setting.
This is an important consistency check, since our method is entirely built on the fact that the environment correlation functions completely determine the dynamics of the open system.
We make sure that, in fact, the open system ``sees'' the expected correlation functions, at least to a sufficient degree of approximation.
To this purpose we choose a non-trivial example, merging the two semicircle spectral densities
\begin{equation}
  \begin{aligned}
    J\Left(\omega)&=\tfrac{1}{2\pi}\sqrt{\omega(2-\omega)},
   &T\Left&=0.2,
   &\mu\Left&=1,\\
    J\Right(\omega)&=\tfrac{1}{4\pi}\sqrt{\omega(2-\omega)},
   &T\Right&=1,
   &\mu\Right&=\tfrac14,
  \end{aligned}
  \label{eq:correlation_function_check_spectral_densities}
\end{equation}
both on the domain \((0,2)\).
Through the procedure detailed in the previous chapters we derive the two equivalent environments
\begin{equation}
  \begin{aligned}
    J^{(0)}(\omega)&=\sum_{\alpha\in\{\Left*,\Right*\}}(1-n_\alpha(\omega))J_\alpha(\omega+\mu_\alpha),\\
    J^{(1)}(\omega)&=\sum_{\alpha\in\{\Left*,\Right*\}}n_\alpha(\omega)J_\alpha(\omega+\mu_\alpha),
  \end{aligned}
  \label{eq:correlation_function_check_equivalent_spectral_densities}
\end{equation}
and reshape the respective environments with the chain mapping: the top half of \cref{fig:correlation_function_check_semicircles} shows how the chain coefficients converge towards their asymptotic values.
We apply the Markovian closure leaving \(N\Environment=13\) or \(N\Environment=20\) chain sites and adding \(N\Closure=6\) pseudomodes on each side: we choose these numbers so that for \(n>N\Environment\) the distance of the \(n\)-th coefficients from their asymptotic value is less than \(10^{-2}\) for \(N\Environment=13\), and less than \(\num{5e-3}\) for \(N\Environment=20\).
We then reproduced the spectral densities by calculating the inverse Fourier transform of the correlation function, which for the initially empty environment gives
\begin{equation}
  \frac{1}{\pi}\Re{}\int_{0}^{+\infty}\dd t\,e^{i\omega t}\eta_0^2\Tr\bigl(c_{0,0}\Phi_t(\adj{c_{0,0}}\ket{\vacuum,\filled}\bra{\vacuum,\filled})\bigr)
  \label{eq:spectral_density_from_simulated_correlation_function}
\end{equation}
where \(c_{0,0}\) is the annihilation operator of the first mode of the chain derived from the initially empty environment, as in~\eqref{eq:cmapExc}, and \(\Phi_t\) the time-evolution map, i.e.~\(\exp(t\lindblad)\) where \(\lindblad\) is the Lindbladian operator fixed by Eqns.~\eqref{eq:hamiltoniano_libero_pseudomodi} and~\eqref{eq:dissipatore_pseudomodi}.
An analogous formula holds for the initially filled environment.
We run the simulation up until \(t=400\); this (inevitable) truncation would introduce artifacts in the Fourier transform, so we also multiply the correlation function by an exponentially decaying factor \(\exp(-at^2)\) with \(a\) such that the resulting product is \(10^{-15}\) at \(t=400\).
This does not alter the correlation function significantly since it is already exponentially decaying, but it smoothes out the errors due to the truncation.
We can see in \cref{fig:correlation_function_check_semicircles} that the expected and simulated spectral densities are in very good agreement.
Of course, since we can compute the correlation function only for a finite amount of physical time, this procedure is not able to describe correctly the spectral density function in a neighborhood of its non-differentiable points; this is expected and, in any case, the deviation is sufficiently small.

\begin{figure}
  \centering
  \begingroup

\pgfplotstableread[col sep=comma]{data_semicircles.thermofield}\coefficients
\pgfplotstableread[col sep=comma, x=frequency]{data_correlation_function_check_semicircles.csv}\data

\pgfmathsetmacro{\vsep}{1.5 cm}
\pgfmathsetmacro{\hsep}{0.6 cm}
\pgfmathsetmacro{\pwidth}{\columnwidth-0.5cm}
\pgfmathsetmacro{\pheight}{\columnwidth-0.5cm}

\pgfdeclarelayer{background}
\pgfdeclarelayer{foreground}
\pgfsetlayers{background,main,foreground}

\begin{tikzpicture}[remember picture]
  \begin{pgfonlayer}{background}
    \begin{groupplot}[
        group style = {
          rows = 2,
          columns = 1,
          vertical sep = \vsep,
        },
        trim axis left,
        xtick align = outside,
        ytick align = outside,
        ylabel near ticks,
        width = \pwidth,
        grid = major,
        enlargelimits,
        no markers, 
        every axis plot/.append style={very thick},
      ]
      \nextgroupplot[
        height=0.75*\pwidth,
        table/x expr=\coordindex,
        xmax=50,
        ymode=log,
        xlabel=$n$,
        legend columns = 2,
        legend style={
          /tikz/every even column/.append style={column sep=0.5cm},
          draw=none,
          font=\small
        }
      ]
      \addplot+ [solid, strongred]   table [y expr=abs(11/16-\thisrow{coupempty})]  \coefficients;
      \addplot+ [solid, strongblue]  table [y expr=abs(11/16-\thisrow{coupfilled})] \coefficients;
      \addplot+ [dotted, strongred]  table [y expr=abs(3/8-\thisrow{freqempty})]    \coefficients;
      \addplot+ [dotted, strongblue] table [y expr=abs(3/8-\thisrow{freqfilled})]   \coefficients;
      %
      \legend{
        {$\abs{\kappa_{0,n}-\homcoupling}$},
        {$\abs{\kappa_{1,n}-\homcoupling}$},
        {$\abs{\omega_{0,n}-\homfrequency}$},
        {$\abs{\omega_{1,n}-\homfrequency}$},
      }
      \nextgroupplot[
        height = \pwidth,
        xlabel = $\omega$,
        xmin = -1.5,
        xmax = 2.3,
        ymin = 0,
        ymax = 0.35,
      ]
      \addplot+ [strongred] table [y expr=0.209/(2*pi)*\thisrow{empty20simulated}] \data;
      \addplot+ [strongred!50] table [y expr=0.209/(2*pi)*\thisrow{empty13simulated}] \data%
      node [strongred, pos=0.7, above right] {\(J^{(0)}(\omega)\)};
      \addplot+ [black, dashed] table [y expr=\thisrow{emptyexpected}] \data;
      \addplot+ [strongblue] table [y expr=0.166/(2*pi)*\thisrow{filled20simulated}] \data;
      \addplot+ [strongblue!50] table [y expr=0.166/(2*pi)*\thisrow{filled13simulated}] \data%
      node [strongblue, pos=0.43, above] {\(J^{(1)}(\omega)\)};
      \addplot+ [black, dashed] table [y expr=\thisrow{filledexpected}] \data;
      \coordinate (inset) at (rel axis cs: 0.95,0.95);
    \end{groupplot}
  \end{pgfonlayer}
  \begin{pgfonlayer}{foreground}
    \begin{semilogyaxis}[
        at={(inset)},
        name = insetaxis,
        axis background/.style={fill=white},
        anchor={outer north east},
        footnotesize,
        xtick align = outside,
        ytick align = outside,
        ylabel near ticks,
        height = 0.4*\pheight,
        width = 0.4*\pwidth+\hsep,
        xmin = -1.5,
        xmax= 2.3,
        grid = major,
        no markers,
        table/x = frequency, 
      ]
      \addplot+ [strongred] table [y expr=abs(\thisrow{emptyexpected}-0.209/(2*pi)*\thisrow{empty20simulated})] \data;
      \addplot+ [strongred!50] table [y expr=abs(\thisrow{emptyexpected}-0.209/(2*pi)*\thisrow{empty13simulated})] \data;

      \addplot+ [strongblue] table [y expr=abs(\thisrow{filledexpected}-0.166/(2*pi)*\thisrow{filled20simulated})] \data;
      \addplot+ [strongblue!50] table [y expr=abs(\thisrow{filledexpected}-0.166/(2*pi)*\thisrow{filled13simulated})] \data;
    \end{semilogyaxis}
  \end{pgfonlayer}
  \begin{pgfonlayer}{main}
    \fill [black!0] ([shift={(-1pt,-1pt)}] insetaxis.outer south west)
      rectangle ([shift={(+5pt,+5pt)}] insetaxis.outer north east);
  \end{pgfonlayer}
\end{tikzpicture}
\endgroup
  \caption{%
    \textit{Top}: absolute distance between chain coefficients of the equivalent environments (\(j=0\): initially empty, \(j=1\): initially filled) obtained from \cref{eq:correlation_function_check_spectral_densities}.
    \textit{Bottom}: comparison of the expected (dashed) and simulated (solid) correlation functions from \cref{eq:correlation_function_check_equivalent_spectral_densities}.
    The inset plot shows the absolute error between the simulated and the expected functions.
    Plots with darker colors represent simulations with \(N\Environment=20\), lighter ones with \(N\Environment=13\).
  }
  \label{fig:correlation_function_check_semicircles}
\end{figure}

\subsection{Accuracy}
\label{ssec:correctness}
We take the non-interacting SIAM with the impurity starting from the occupied state, while the environment is described by a semicircle spectral density
\begin{equation}
  J(\omega) = \frac{1}{10\pi} \sqrt{\omega(2-\omega)}
  \label{eq:benchmark_spectral_density}
\end{equation}
on the domain \((0,2)\), and \(\epsilon=-\tfrac{\pi}{8}\) as the energy of the excited level of the impurity.

\begin{figure}
  \centering
  \begingroup
\pgfmathsetmacro{\vsep}{1 cm}
\pgfmathsetmacro{\hsep}{0.6 cm}
\pgfmathsetmacro{\pwidth}{\columnwidth-0.5cm}
\pgfmathsetmacro{\pheight}{\columnwidth-0.5cm}
\pgfplotstableread{data_siam_mu0.2_pure40_measurements.dat}\tedopa%
\pgfplotstableread{data_siam_mu0.2_NE8_mc100_NC6_measurements.dat}\mca%
\pgfplotstableread{data_siam_mu0.2_NE8_mc100_NC8_measurements.dat}\mcb%
\pgfplotstableread{data_siam_mu0.2_NE8_mc100_NC10_measurements.dat}\mcc%
\pgfplotstablecreatecol[copy column from table={\tedopa}{n{1}_re}]{ref}{\mca}%
\pgfplotstablecreatecol[copy column from table={\tedopa}{n{1}_re}]{ref}{\mcb}%
\pgfplotstablecreatecol[copy column from table={\tedopa}{n{1}_re}]{ref}{\mcc}%

\pgfdeclarelayer{background}
\pgfdeclarelayer{foreground}
\pgfsetlayers{background,main,foreground}

\begin{tikzpicture}[remember picture]
  \begin{pgfonlayer}{background}
    \begin{axis}[
        xtick align = outside,
        ytick align = outside,
        ylabel near ticks,
        xlabel     = \(t\),
        ylabel     = \(\avg{n\System(t)}\),
        legend style = {
          font = \footnotesize,
          draw = none,
          at = {(rel axis cs:0.95, 0.05)},
          anchor = south east,
          /tikz/every even column/.append style={column sep=0.5cm},
        },
        legend columns = 2,
        height     = \pheight,
        width      = \pwidth,
        grid       = major,
        cycle list name = short,
        table/x = time, 
        ymin = 0.885,
      ]
      \addplot+ table[y expr = \thisrow{vN{1}_re}/\thisrow{Norm}] \mca;
      \addplot+ table[y expr = \thisrow{vN{1}_re}/\thisrow{Norm}] \mcb;
      \addplot+ table[y expr = \thisrow{vN{1}_re}/\thisrow{Norm}] \mcc;
      \addplot[thick, dashed, black] table[y expr=\thisrow{n{1}_re}] \tedopa;
      \legend{
        \(N\Closure=6\),
        \(N\Closure=8\),
        \(N\Closure=10\),
        TEDOPA
      }
      \coordinate (inset) at (rel axis cs: 0.95,0.95);
    \end{axis}
  \end{pgfonlayer}
  %
  \begin{pgfonlayer}{foreground}
    \begin{semilogyaxis}[
        at={(inset)},
        name = insetaxis,
        axis background/.style={fill=white},
        anchor={outer north east},
        footnotesize,
        xtick align = outside,
        ytick align = outside,
        ylabel near ticks,
        height     = 0.5*\pheight,
        width      = 0.5*\pwidth+\hsep,
        grid       = major,
        cycle list name = short,
        table/x = time, 
        table/y expr = abs(\thisrow{vN{1}_re}/\thisrow{Norm}-\thisrow{ref})
      ]
      \addplot+ table \mca;
      \addplot+ table \mcb;
      \addplot+ table \mcc;
    \end{semilogyaxis}
  \end{pgfonlayer}

  \begin{pgfonlayer}{main}
    \fill [black!0] ([shift={(-1pt,-1pt)}] insetaxis.outer south west)
    rectangle ([shift={(+5pt,+5pt)}] insetaxis.outer north east);
  \end{pgfonlayer}
\end{tikzpicture}
\endgroup
  \caption{%
    Population of the open system, starting from a filled state, comparing the FMC and standard TEDOPA methods on a spinless SIAM simulation with \(\Omega=(-\num{0.2},\num{1.8})\), \(\mu=0\) and \(T=\num{0.4}\), \(N\Environment=8\), \(\epsilon=-\tfrac{\pi}{8}\).
    Inset: absolute error between FMC and TEDOPA simulations.%
  }
  \label{fig:benchmark_spinless_siam_mu0.2_NE8}
\end{figure}

\begin{figure}
  \centering
  \begingroup
\pgfmathsetmacro{\vsep}{1 cm}
\pgfmathsetmacro{\hsep}{0.6 cm}
\pgfmathsetmacro{\pwidth}{\columnwidth-0.5cm}
\pgfmathsetmacro{\pheight}{\columnwidth-0.5cm}
\pgfplotstableread{data_siam_mu1_p20_measurements.dat}\tedopa%
\pgfplotstableread{data_siam_mu1_NE6_mc200_NC6_measurements.dat} \mca%
\pgfplotstableread{data_siam_mu1_NE6_mc220_NC8_measurements.dat} \mcb%
\pgfplotstableread{data_siam_mu1_NE6_mc240_NC10_measurements.dat}\mcc%
\pgfplotstablecreatecol[copy column from table={\tedopa}{n{1}_re}]{ref}{\mca}%
\pgfplotstablecreatecol[copy column from table={\tedopa}{n{1}_re}]{ref}{\mcb}%
\pgfplotstablecreatecol[copy column from table={\tedopa}{n{1}_re}]{ref}{\mcc}%
\pgfdeclarelayer{background}
\pgfdeclarelayer{foreground}
\pgfsetlayers{background,main,foreground}

\begin{tikzpicture}[remember picture]
  \begin{pgfonlayer}{background}
    \begin{axis}[
        xtick align = outside,
        ytick align = outside,
        ylabel near ticks,
        xlabel     = \(t\),
        ylabel     = \(\avg{n\System(t)}\),
        legend style = {
          font = \footnotesize,
          draw = none,
          at = {(rel axis cs:0.95, 0.25)},
          anchor = east,
          /tikz/every even column/.append style={column sep=0.5cm},
        },
        legend columns = 2,
        height     = \pheight,
        width      = \pwidth,
        grid       = major,
        cycle list name = short,
        table/x = time,
        table/y expr = \thisrow{vN{1}_re}/\thisrow{Norm} 
      ]
      \addplot+ table \mca;
      \addplot+ table \mcb;
      \addplot+ table \mcc;
      \addplot[thick, dashed, black] table[y expr=\thisrow{n{1}_re}] \tedopa;
      \legend{
        \(N\Closure=6\),
        \(N\Closure=8\),
        \(N\Closure=10\),
        TEDOPA
      }
      \coordinate (inset) at (rel axis cs: 0.95,0.95);
    \end{axis}
  \end{pgfonlayer}
  \begin{pgfonlayer}{foreground}
    \begin{semilogyaxis}[
        at={(inset)},
        name = insetaxis,
        axis background/.style={fill=white},
        anchor={outer north east},
        footnotesize,
        xtick align = outside,
        ytick align = outside,
        ylabel near ticks,
        height     = 0.5*\pheight,
        width      = 0.5*\pwidth+\hsep,
        grid       = major,
        cycle list name = short,
        table/x = time, 
        table/y expr = abs(\thisrow{vN{1}_re}/\thisrow{Norm}-\thisrow{ref})
      ]
      \addplot+ table \mca;
      \addplot+ table \mcb;
      \addplot+ table \mcc;
    \end{semilogyaxis}
  \end{pgfonlayer}

  \begin{pgfonlayer}{main}
    \fill [black!0] ([shift={(-1pt,-1pt)}] insetaxis.outer south west)
      rectangle ([shift={(+5pt,+5pt)}] insetaxis.outer north east);
  \end{pgfonlayer}
\end{tikzpicture}
\endgroup
  \caption{%
    Population of the open system, starting from a filled state, comparing the FMC and standard TEDOPA methods on a spinless SIAM simulation with \(\Omega=(-1,1)\), \(\mu=0\), \(T=\num{0.4}\) and \(N\Environment=6\), \(\epsilon=-\tfrac{\pi}{8}\).
    Inset: absolute error between FMC and TEDOPA simulations.%
  }
  \label{fig:benchmark_spinless_siam_mu1}
\end{figure}

In \cref{fig:benchmark_spinless_siam_mu0.2_NE8,fig:benchmark_spinless_siam_mu1} we compare our approach to standard TEDOPA, i.e.~without the Markovian closure, by plotting the expectation value of the population of the impurity site.
We also translate \(J\) so that the chemical potential is zero, obtaining a new frequency domain \(\Omega=(-\num{0.2},\num{1.8})\) from \(\mu=0.2\) in \cref{fig:benchmark_spinless_siam_mu0.2_NE8} and \(\Omega=(-1,1)\) from \(\mu=1\) in \cref{fig:benchmark_spinless_siam_mu1}.
In both figures different closure sizes (\(N\Closure=6,8,10\)) attached to the truncation point \(N\Environment=6\) are considered.

The number \(N\Closure\) of pseudomodes has an effect on the accuracy of the simulations.
Counter-intuitively, the use of a larger number of pseudomodes does not necessarily improve the quality of the results.
As observed in Ref.~\onlinecite{Nuesseler2022:markovian_closure}, this can be related to the fact that the \(N\Closure=8\) closure better approximates the residual spectral density near the border of the domain than the \(N\Closure=10\) one.
The quality of the approximation of the semicircle spectral density at the edges of the support impacts on low- and high-momentum components of the wave-packet traveling along the chain.
The effect of the deviations from the ideal spectral density in these regions, therefore, strongly depends on the excitation dynamics on the chain, which is, in turn, determined by the chain coefficients.
Unbalanced environments, namely when the shifted support $\Omega-\mu$ of the spectral density $J_{\beta,\mu}(\omega)$ is not symmetric with respect to the origin, will lead to different system-chain coupling strength $\eta_j$ (see \cref{eq:cmapExc}) for the vacuum ($j=0$) and filled chain ($j=1$).
This asymmetry is responsible for the generation of slowly traveling packets, which, due to the chain momentum/energy dispersion relation, sample the frequency region where the approximation of the asymptotic spectral density is less accurate.
If the chemical potential sits at the middle point of the domain of the spectral density, as in the case considered in \cref{fig:benchmark_spinless_siam_mu1}, the situation is more favorable: the system-chain couplings $\eta_j$ have the same magnitude and the Markovian closure performs better, without a significant difference between the three \(N\Closure=6,8,10\) cases.

The accuracy does not depend, on the other side, on \(N\Environment\), i.e.~how many sites we leave on the original chains before we attach the FMC; if we increase this parameter, the error starts to appear at a later time (as expected), but then it increases until it reaches the same value.
We show in \cref{sec:convergence_chain_coefficients_from_simulations} that with our choices of \(N\Environment\) we can reproduce the spectral density functions of the equivalent environment to a very high accuracy.
Increasing \(N\Environment\) also increases the numerical costs in general, since the system is bigger, and possibly the bond dimension too, so we see no practical advantage in taking a bigger value of this parameter.

Simulations of spinful SIAM are inevitably more costly than the spinless counterpart, since the local dimension increases from \(2^2=4\) to \(4^2=16\), merging the two spin components into a single vector in \(\C^4\).
In this case, as discussed in Ref.~\onlinecite{Nuesseler2020:ftedopa}, a simulation in the Heisenberg picture, which involves minimal modifications to the time-evolution algorithm in our vectorized setting, allows us to use an overall lower bond dimension.
\Cref{fig:spinful_siam_Ntot_system} compares the results in the spinful obtained by Heisenberg-picture standard TEDOPA and TEDOPA+FMC, with \cref{eq:benchmark_spectral_density} as spectral density, \(\epsilon_{\spinup}=\epsilon_{\spindown}=-\tfrac{\pi}{8}\) and \(U=-\tfrac{\pi}{4}\), \(T=\num{0.4}\) and \(\mu=1\).
We observe that after a reasonably long initial transient period the FMC has introduced an error of the order of \(10^{-3}\) in all cases, both spinless and spinful ones.

\begin{figure}
  \centering
  \begingroup
\pgfmathsetmacro{\vsep}{1 cm}
\pgfmathsetmacro{\hsep}{0.6 cm}
\pgfmathsetmacro{\pwidth}{\columnwidth-0.5cm}
\pgfmathsetmacro{\pheight}{\columnwidth-0.5cm}
\newcommand\tedopacolname{(p100)Ntot{1}_re}
\newcommand\mccolname{(mc150_H)Ntot{1}_re}
\pgfplotstableread[col sep=comma]{data_siam_mu1_spinful_p100+mc150_H_measurements.csv}\data
\pgfdeclarelayer{background}
\pgfdeclarelayer{foreground}
\pgfsetlayers{background,main,foreground}

\begin{tikzpicture}[remember picture]
  \begin{pgfonlayer}{background}
    \begin{axis}[
        xtick align = outside,
        ytick align = outside,
        ylabel near ticks,
        xlabel     = \(t\),
        ylabel     = \(\avg{n_{\System*,\uparrow}(t)+n_{\System*,\downarrow}(t)}\),
        legend style = {
          draw = none,
          font = \footnotesize,
          at = {(rel axis cs:0.95, 0.75)},
          anchor = east,
        },
        height     = \pheight,
        width      = \pwidth,
        grid       = major,
        cycle list name = short,
        table/x = time, 
      ]
      \pgfplotsset{cycle list shift=1}
      \addplot+ table[y expr = \thisrow{\mccolname}] \data;
      \addplot[thick, dashed, black] table[y expr=\thisrow{\tedopacolname}] \data;
      \legend{
        \(N\Closure=8\),
        TEDOPA
      }
      \coordinate (inset) at (rel axis cs: 0.95,0.05);
    \end{axis}
  \end{pgfonlayer}
  \begin{pgfonlayer}{foreground}
    \begin{semilogyaxis}[
        at={(inset)},
        name = insetaxis,
        axis background/.style={fill=white},
        anchor={outer south east},
        footnotesize,
        xtick align = outside,
        ytick align = outside,
        ylabel near ticks,
        height     = 0.5*\pheight,
        width      = 0.5*\pwidth+\hsep,
        grid       = major,
        cycle list name = short,
        table/x = time, 
        table/y expr = abs(\thisrow{\mccolname}-\thisrow{\tedopacolname})
      ]
      \pgfplotsset{cycle list shift=1}
      \addplot+ table \data;
    \end{semilogyaxis}
  \end{pgfonlayer}

  \begin{pgfonlayer}{main}
    \fill [black!0] ([shift={(-1pt,-1pt)}] insetaxis.outer south west)
      rectangle ([shift={(+5pt,+5pt)}] insetaxis.outer north east);
  \end{pgfonlayer}
\end{tikzpicture}
\endgroup
  \caption{Expectation value of the total population \(\avg{n_{\System*,\spinup}(t)+n_{\System*,\spindown}(t)}\) of the system site in the spinful SIAM: we show an FMC simulation in the Heisenberg picture, with \(J\) as in \cref{eq:benchmark_spectral_density}, \(\epsilon_{\spinup}=\epsilon_{\spindown}=-\tfrac{\pi}{8}\), \(U=-2\epsilon_{\spinup}\), \(T=\num{0.4}\) and \(\mu=1\), and a standard TEDOPA one taken as a reference. Inset: absolute error between FMC and TEDOPA simulations.}
  \label{fig:spinful_siam_Ntot_system}
\end{figure}

\subsection{Computational cost}
For a given chain of length $L$, the complexity of a TEDOPA simulation scales as \(O(Lt\Max (d\chi)^3)\)~\cite{Tamascelli2019:ttedopa}, where \(d\) is the local dimension, \(\chi\) the bond dimension of the MPS, and \(t\Max\) the total physical simulation time.
In a uniform TEDOPA chain, the excitations propagate at rate that equals twice the coupling constant between the sites.
In a typical TEDOPA simulation the asymptotic value of the coupling constant, \(K\), is a good estimate of the actual propagation speed if the coefficient, site by site, converges relatively quickly towards this value.
The chain should therefore contain at least \(2K t\Max\) sites if we want to avoid that the excitations ``bounce back'', creating artifacts due to the finite size; with \(L=K t\Max\) we get that TEDOPA complexity scales as \(O(K t\Max^2d^3\chi^3)\).

In FMC simulations, the size of the system (i.e.~its length \(L\)) is fixed, but \(d\) gets squared and the new bond dimension \(\chi'\) is generally higher than \(\chi\), to account for thermal correlations and the presence of longer-range interactions: overall we find \(O(t\Max d^6\chi^{\prime3})\).
Whether the FMC is an advantage over standard TEDOPA depends on how much \(\chi'\) needs to be higher than \(\chi\).

Given these scaling properties, it is clear that the FMC can be an advantage over a standard TEDOPA simulation only if we want to study long-time dynamics, when the \(t\Max^2\)-dependence of the latter can overcome the simpler \(t\Max\)-dependence of the former.
In \cref{ssec:correctness} we compared the two methods to prove that the FMC provides a satisfactory good approximation to a standard TEDOPA simulation.
To compare directly the two methods in terms of efficiency would however be unfair, since they are really meant to be used in different situations:
\begin{itemize}
  \item a standard TEDOPA simulation is more efficient, even with very long chains, if the entanglement is low, effectively making use of the favorable scaling properties of matrix-product states;
  \item the FMC method works best with very long simulations or if the whole system develops moderate levels of entanglement, in which case the determination of the evolution of long chains would require very long computational times.
\end{itemize}
It is however important to stress that, if the open system starts from a mixed state or if there are Lindblad terms acting on the open system (e.g.~dephasing terms), so that both TEDOPA and FMC simulation methods have to deal with density matrices and thus a \(d^6\)-dependence on the local dimension, then the FMC provides a clear major speed-up, not limited to long time simulations, due to its naturally finite size.

The possibility to reach longer physical times can be beneficial e.g.~for the computation of spectral functions: usually, extrapolation of the retarded or advanced Green functions by means of some ingenious method, such as linear prediction~\cite{ganahl15}, is needed.
The determination of the retarded Green function typically requires the preparation of some equilibrium state through an adiabatic evolution from a simple initial state.
This evolution must be sufficiently slow so as to reach the correct equilibrium state, so that one can easily end up with very long TEDOPA chains in order to accommodate a high \(t\Max\).
We try out this scheme on the non-interacting model, following the technique presented in Ref.~\onlinecite{Kohn2021:interleaved_mapping}: system and environment are initialized in a product state, then they are evolved using a time-dependent Hamiltonian where the system-environment interaction term is slowly ramped up from zero to its actual value.
\Cref{fig:equilibration_spinless_siam_mu1} displays the results of this procedure: as always, we take a standard TEDOPA simulation as a reference.

\begin{figure}
  \centering
  \begingroup
\pgfmathsetmacro{\vsep}{5 mm}
\pgfmathsetmacro{\pwidth}{\columnwidth-0.5cm}
\pgfmathsetmacro{\pheight}{0.8*\pwidth}
\pgfplotstableread{data_equilibration_siam_spinless_mu1_slope0.05_p30_measurements.dat}\tedopa%
\pgfplotstableread{data_equilibration_siam_spinless_mu1_slope0.05_mc200_NC6_measurements.dat}\mcb%
\pgfplotstableread{data_equilibration_siam_spinless_mu1_slope0.05_mc200_NC8_measurements.dat}\mcc%
\pgfplotstablecreatecol[copy column from table={\tedopa}{n{1}_re}]{ref1}{\mcb}%
\pgfplotstablecreatecol[copy column from table={\tedopa}{n{1}_re}]{ref1}{\mcc}%
%
\pgfplotstablecreatecol[copy column from table={\tedopa}{n{11}_re}]{ref11}{\mcb}%
\pgfplotstablecreatecol[copy column from table={\tedopa}{n{11}_re}]{ref11}{\mcc}%
\pgfdeclarelayer{background}
\pgfdeclarelayer{foreground}
\pgfsetlayers{background,main,foreground}
\begin{tikzpicture}[remember picture]
  \begin{pgfonlayer}{background}
    \begin{groupplot}[
        group style = {
          columns = 1,
          rows = 2,
          vertical sep = \vsep,
          x descriptions at = edge bottom,
        },
        xtick align = outside,
        ytick align = outside,
        ylabel near ticks,
        height = \pheight,
        width = \pwidth,
        grid = major,
        cycle list name = short,
        table/x = time, 
        xlabel = \(t\),
      ]
      \nextgroupplot[
        ylabel = $\avg{n\System(t)}$,
        legend style = {
          draw = none,
          font = \footnotesize,
          at = {(rel axis cs:0.95, 0.725)},
          anchor = east,
          /tikz/every even column/.append style={column sep=0.5cm},
        },
        legend columns = 1,
      ]
      \addplot+ table[y expr = \thisrow{vN{1}_re}/\thisrow{Norm}] \mcb;
      \addplot+ table[y expr = \thisrow{vN{1}_re}/\thisrow{Norm}] \mcc;
      \addplot[thick, dashed, black] table[y expr=\thisrow{n{1}_re}] \tedopa;
      \legend{
        {\(N\Closure=6\)},
        {\(N\Closure=8\)},
        TEDOPA
      }
      \coordinate (inset1) at (rel axis cs: 0.95,0.05);
      \nextgroupplot[
        ylabel = \(\avg{n_{\Environment*,1|N\Environment}(t)}\),
        ymax = 1.0195,
      ]
      \addplot+ table[y expr = \thisrow{vN{11}_re}/\thisrow{Norm}] \mcb;
      \addplot+ table[y expr = \thisrow{vN{11}_re}/\thisrow{Norm}] \mcc;
      \addplot[thick, dashed, black] table[y expr=\thisrow{n{11}_re}] \tedopa;
      \coordinate (inset11) at (rel axis cs: 0.95,0.95);
    \end{groupplot}
  \end{pgfonlayer}
  \begin{pgfonlayer}{foreground}
    \begin{semilogyaxis}[
        at={(inset1)},
        name = inset1axis,
        axis background/.style={fill=white},
        anchor={outer south east},
        footnotesize,
        xtick align = outside,
        ytick align = outside,
        ylabel near ticks,
        height     = 0.5*\pheight,
        width      = 0.5*\pwidth,
        grid       = major,
        cycle list name = short,
        table/x = time, 
        table/y expr = abs(\thisrow{vN{1}_re}/\thisrow{Norm}-\thisrow{ref1})
      ]
      \addplot+ table \mcb;
      \addplot+ table \mcc;
    \end{semilogyaxis}
    \begin{semilogyaxis}[
        at={(inset11)},
        name = inset11axis,
        axis background/.style={fill=white},
        anchor={outer north east},
        footnotesize,
        xtick align = outside,
        ytick align = outside,
        ylabel near ticks,
        height     = 0.5*\pheight,
        width      = 0.5*\pwidth,
        grid       = major,
        cycle list name = short,
        table/x = time, 
        table/y expr = abs(\thisrow{vN{11}_re}/\thisrow{Norm}-\thisrow{ref11})
      ]
      \addplot+ table \mcb;
      \addplot+ table \mcc;
    \end{semilogyaxis}
  \end{pgfonlayer}
  \begin{pgfonlayer}{main}
    \fill [black!0] ([shift={(-1pt,-1pt)}] inset1axis.outer south west)
        rectangle ([shift={(+5pt,+5pt)}] inset1axis.outer north east);
    \fill [black!0] ([shift={(-1pt,-1pt)}] inset11axis.outer south west)
        rectangle ([shift={(+5pt,+5pt)}] inset11axis.outer north east);
  \end{pgfonlayer}
\end{tikzpicture}
\endgroup
  \caption{%
    Adiabatic evolution of non-interacting SIAM with \(T=\num{0.4}\), \(\mu=1\), \(\epsilon=-\frac{\pi}{8}\) and \(J\) given by \cref{eq:benchmark_spectral_density}.
    Population of the open system (top frame) and of the last site before the FMC in the chain of the initially filled environment (bottom frame). 
    The composite system is evolved from a product state towards a state where the impurity is in equilibrium with the environment.
    We show two different approximations within the FMC scheme, varying the number \(N\Closure\) of pseudomodes.
    Insets: absolute error between the FMC and TEDOPA simulations.
  }
  \label{fig:equilibration_spinless_siam_mu1}
\end{figure}

We use the spectral density \(J\) as in \cref{eq:benchmark_spectral_density} and \(\epsilon=-\tfrac{\pi}{8}\).
At the beginning of the evolution, our open system is in the empty state and the environment in its thermal equilibrium state with \(T=\num{0.4}\) and \(\mu=1\); the composite system is slowly evolved under the Hamiltonian \(H(t)=H\Environment+r(t)(H\Interaction+H\System)\) where \(r(t)=\min\{1,t/\tau\}\), \(\tau=20\), until \(t=100\).
We found that a bond dimension equal to \num{200} is sufficient for an FMC structure with \(N\Environment=6\) and \(N\Closure=6\) to obtain a state \(\rho_{\textnormal{MC}}\) such that, if \(\rho_{\textnormal{TEDOPA}}\) is the state obtained with a standard TEDOPA evolution, \(\abs{\Tr(A(\rho_{\textnormal{MC}}-\rho_{\textnormal{TEDOPA}}))}<10^{-3}\) for local and non-local observables \(A\) acting on the common sites between the two systems, i.e.~on the open system and on the first \(N\Environment\) sites on each environment chain.
We show at last in \cref{fig:walltime} the difference in computational time between the standard TEDOPA and FMC simulations used to compute these equilibrium states.
It illustrates the wall-clock time required for each time step in the relaxation phase (i.e.~when \(t>\tau\)) of the adiabatic evolution towards an equilibrium state, considering environments of various lengths.
In this situation, entanglement is slowly spreading along the chain, requiring more and more time for each sweep of the TDVP algorithm in the normal TEDOPA cases, since the number of singular values needed to well approximate the state increases and so the singular value decomposition~(SVD) takes more time.
In the FMC case, instead, the entanglement increases rapidly at the beginning (which here already happened in the ramp-up phase of the adiabatic evolution, not shown) and reaches very soon the maximum quantity allowed by the limited bond dimension.
After this point, the computational time remains constant as the evolution goes on.
\begin{figure}
  \centering
  \begingroup
\pgfmathsetmacro{\vsep}{1 cm}
\pgfmathsetmacro{\hsep}{0.6 cm}
\pgfmathsetmacro{\pwidth}{\columnwidth-0.5cm}
\pgfmathsetmacro{\pheight}{\columnwidth-0.5cm}
\pgfplotstableread[col sep=comma]{data_walltime_pure_relax_cropped.csv}\data%
\pgfplotstableread{data_walltime_mc.dat}\datamc%
\pgfplotsset{unit code/.code 2 args={\si{#1#2}}}
\pgfplotscreateplotcyclelist{pairedshort}{
    {strongblue!40},
    {strongblue},
    {stronggreen!40},
    {stronggreen},
    {strongyellow!40},
    {strongyellow},
    {strongred!40},
    {strongred},
    {stronggray!40},
    {stronggray}
}
\begin{tikzpicture}
  \begin{groupplot}[
      group style = {
        columns = 1,
        rows = 2,
        vertical sep = 1.75*\vsep,
      },
      xtick align = outside,
      ytick align = outside,
      ylabel near ticks,
      xlabel = Step,
      height = \pheight,
      width = \pwidth,
      grid = major,
      cycle list name = pairedshort,
      unit markings = slash space,
    ]
    \nextgroupplot[
      ylabel = Step wall-clock time,
      y unit = \second,
      forget plot style={opacity=0.3},
      xmin = 0,
      xmax = 150,
      table/x expr = \coordindex,
    ]
    \pgfplotsinvokeforeach{60, 70, 80, 90, 100, 110, 120, 130} {%
      \addplot+ table[y=L#1] \data;
      \addplot+[thick] table[y={create col/linear regression={y=L#1}}] {\data} node[pos=0.9, sloped, below] {#1};
    }
    \addplot[forget plot, black] table[col sep=comma, x expr=2*\coordindex, y=mc200ne6nc6] \datamc;
    \addplot[thick, black] table[col sep=comma, x expr=2*\coordindex, y={create col/linear regression={y=mc200ne6nc6}}] {\datamc} node[pos=0.42, sloped, below] {FMC};
    \nextgroupplot[
      table/col sep = comma,
      xlabel = $t\Max$,
      ylabel = Total wall-clock time,
      y unit = \second,
      legend style = {
        draw = none,
        font = \small,
        at = {(rel axis cs:0.05, 0.95)},
        anchor = north west,
        /tikz/every even column/.append style={column sep=0.5cm},
      },
      enlarge x limits,
      xmin = 60,
      xmax = 140,
      height = 0.75*\pheight,
      width = \pwidth,
    ]
    \addplot[strongred, mark=x, thick, only marks, mark size=4pt] table {data_walltime_total.csv};
    \addplot[strongblue, domain=0:200] {2 * 24.262113 * x};
    \legend{TEDOPA,FMC}
  \end{groupplot}
\end{tikzpicture}
\endgroup
  \caption{%
    Computational time required for each step (top) and for the whole evolution (bottom) in standard TEDOPA and FMC simulations, for varying lengths of the TEDOPA chain, of the same system in \cref{fig:equilibration_spinless_siam_mu1}.
    All simulations were run with 8~CPU cores on identical hardware.
  }
  \label{fig:walltime}
\end{figure}

\subsection{Possible improvements}
\newcommand{\ndiff}{\mathcal{N}}%
We note that the structure given by the Markovian closure construction resembles the one presented in Ref.~\onlinecite{lotem:renormalized_lindblad_driving}; in fact, some remarks given by the authors in that paper apply here as well.
Specifically, consider the total number operator, which e.g.~for the spinless SIAM in a thermal environment (we use the same \(N\Environment\) and \(N\Closure\) for the two equivalent zero-temperature environments for simplicity)
\begin{equation}
  N\!=
  \adj{d}d
  +\sum_{n=0}^{N\Environment}(\adj{c_{0,n}}c_{0,n}\phantomadj + \adj{c_{1,n}}c_{1,n}\phantomadj)
  +\sum_{n=1}^{N\Closure}(\adj{a_{0,n}}a_{0,n}\phantomadj + \adj{a_{1,n}}a_{1,n}\phantomadj)
\end{equation}
where \(c\) denotes a chain operator as in \cref{eq:approxHam} and \(a\) a closure operator such as in \cref{eq:hamiltoniano_libero_pseudomodi}.
the linear map \(\ndiff(\rho)\defeq\bcomm{N,\rho}\) commutes with the generator \(\lindblad\) of the time evolution, i.e.~\(\lindblad(\ndiff(\rho))=\ndiff(\lindblad(\rho))\), as long as the original Hamiltonian conserves the particle number~\cite{albert:conserved_quantities_lindblad}.

This could be used to restrict the evolution of the initial state within its initial \(\ndiff\)-eigenspace: if the system \(\System*\) starts from a pure state which is an eigenstate of the number operator then \(\ndiff(\rho^0\System\rho^0\Environment)=0\).

Another improvement comes from using a different factorization technique for the re-orthogonalization of the MPS during a TDVP sweep, where the cost of a traditional SVD decomposition can become very demanding (it scales as \((d\chi)^3\) where \(d\) is the local dimension and \(\chi\) the bond dimension) especially in the spinful case where \(d=16\).
The \emph{reduced-rank randomized singular value decomposition}~\cite{tama15} has a more favorable scaling with respect to the standard SVD, and might be useful to speed up the calculations.
{We finally observe that, for very long simulation times that exceed the memory time of the environment, the transfer tensors formalism~\cite{Rosenbach2016:transfer_tensors} can be used to further enhance efficiency.}

\section{Conclusions}
In this work we presented a way to improve the numerical simulation with tensor networks of open systems coupled to continuous fermionic environments.
Firstly, we showed a way to reshape the environments: through the TCSM transformation, which is closely related to the thermofield transformation and absorbs the factors depending on chemical potential and temperature into a modified spectral density, it becomes possible to merge several environments into a single effective one.
Besides a reduction of the number of chains to be determined by means of the TEDOPA chain mapping, this simplification can benefit other simulation schemes. 
Secondly, after performing the chain mapping on the environment to transform it into a linear discrete chain of sites, we developed a way to approximate the dissipative behavior of the environments using only a finite number of sites, extending the already existing MC construction for bosonic environments to fermionic environments.
We compared the time evolution with normal TEDOPA and with the FMC of several systems and environments and found in all cases a satisfactory match between the expectation values of operators measured in the two cases, with relative errors below $10^{-3}$.
Moreover, we successfully compared the environment correlation functions (and the associated spectral density functions) from FMC simulations against the expected theoretical results in some trivial and non-trivial cases.
These results, starting from the equivalence theorem, show the validity of the FMC method.

The FMC can in some cases reduce the time complexity of the simulation and proves to be more efficient than standard TEDOPA in simulations involving long chains and moderate levels of entanglement.
We have therefore developed a general framework that complements TEDOPA in the study of continuous fermionic environments, with a systematic way to approximate the behavior of environments with a finite number of sites that does not rely on specific forms of the spectral density or ad-hoc discretization schemes.
We moreover remark that the FMC, by playing the role of a (quasi-)perfect absorber or emitter, can find application in a variety of cases, independently of the use of chain mapping techniques~\cite{Vittmann:2023aa}.

There are of course some improvements that can be studied in order to enhance the performance of the FMC method.
While the quality of the results obtained by means of FMC is already more than satisfying, we can look for a better parameterization of the closure, e.g.~by means of the techniques introduced in Refs.~\onlinecite{feist21,feist21nano}, to improve the accuracy of the fitting of the asymptotic spectral density, as well as fermion-to-spin representations leading to a decrease of the entanglement in the tensor network.
This will all be object of future work, together with the exploitation of symmetries and conserved quantities leading to a further reduction of the time/memory complexity of the simulation, and the application of this method to analyze physically relevant quantities such as spectral functions of interacting systems or the Kondo peak.

\begin{acknowledgments}
We thank Nicola Lorenzoni, Marco Genoni, Alex W.~Chin, Stephen Clark and Matam Lotem for helpful 
discussions during the completion of this work.
DT and DF were supported by the PSR initiative of the University of Milan.
AS was supported by Ministero dell'Università e della Ricerca under the ``PON Ricerca e Innovazione 2014--2020'' project EEQU and under the ``PRIN 2022'' project PRIN22-2022FEXLYB-QURECO.
MBP and SFH were supported by the ERC Synergy grant HyperQ (grant no.~856432), the QuantERA project ExtraQt (DFG grant no.~499241080) and the EU Project SPINUS (grant no.~101135699).
\end{acknowledgments}

\section*{Code availability}
\begin{myverbbox}{\github}github.com/phaerrax/markovian_closure_fermions\end{myverbbox}%
Data presented in this article were obtained using the ITensor library~\cite{ITensor} in the \emph{Julia} programming language.
All code is available in the GitHub repository \scalebox{.87}[.95]{\href{https://github.com/phaerrax/markovian_closure_fermions}{\github}}.

\appendix

\NewDocumentCommand \ExtEnvironment { s }{%
  \IfBooleanTF#1 {\textnormal{E}'} {_{\textnormal{E}'}}%
}
\section{Thermofield transformation}
\label{sec:thermofield}
We observe that the state $\rho_{\beta,\mu}$ of the environmental fermionic degrees of freedom is, in general, a mixed state. As shown in Refs.~\onlinecite{deVega15, Nuesseler2020:ftedopa}, the same TTCFs, and therefore the same reduced dynamics of the open quantum system $\System*$, is determined by an extended environment $\ExtEnvironment*$ obtained from the original one via a \emph{thermofield transformation}~\cite{taka75} and starting from a pure state.
The transformation is essentially based on a local purification of the mixed state $\rho_{\beta,\mu}$: to each ``physical'' mode $f_{1,\omega} \equiv f_\omega$ we associate an ancillary mode $f_{2,\omega}$, so that the free Hamiltonian of the extended environment is
\begin{equation}\label{eq:extHam}
  H\ExtEnvironment=
  \int_0^{\omega\Max}\dd\omega \, (\omega-\mu)(
    \adj{f_{1,\omega}}f_{1,\omega}\phantomadj -
    \adj{f_{2,\omega}}f_{2,\omega}\phantomadj
  ).
\end{equation}
The ancillary modes are introduced as independent modes, not interacting with the physical ones.
The $f_{1,\omega}$ and $f_{2,\omega}$ modes are then linearly combined into new fermionic modes $c_{1,\omega}$ and $c_{2,\omega}$ through the unitary (orthogonal) transformation~\cite{schwarz18}
\begin{equation}
  \begin{pmatrix}
    c_{1,\omega} \\
    c_{2,\omega}
  \end{pmatrix}
  =
  \begin{pmatrix}
    \cos\bigl(\theta_{\beta,\mu}(\omega)\bigr) &  -\sin\bigl(\theta_{\beta,\mu}(\omega)\bigr) \\
    \sin\bigl(\theta_{\beta,\mu}(\omega)\bigr) & \cos\bigl(\theta_{\beta,\mu}(\omega)\bigr)
  \end{pmatrix}
  \begin{pmatrix}
    f_{1,\omega} \\
    \adj{f_{2,\omega}}
  \end{pmatrix},
\end{equation}
with $\theta_{\beta,\mu}(\omega)$ determined through the relation
\begin{equation} \label{eq:occRel}
  \sin^2\bigl(\theta_{\beta,\mu}(\omega)\bigr)=
  \frac{1}{e^{\beta(\omega-\mu)}+1}.
\end{equation}
In terms of the newly defined fermionic modes $c_{1/2,\omega}$, the free Hamiltonian \eqref{eq:extHam} reads
\begin{equation}
  H_{\Environment*'}=
  \int_0^{\omega\Max} \dd\omega\,(\omega-\mu) (
    \adj{c_{1,\omega}} c_{1,\omega}\phantomadj+
    \adj{c_{2,\omega}} c_{2,\omega}\phantomadj
  ).
\end{equation}

We observe that if the modes $c_{1,\omega}$ start from the (factorized) vacuum state $\ket{\vacuum_1}$ such that $c_{1,\omega}\ket{\vacuum_1} = 0 $ and the modes $c_{2,\omega}$ start from the filled state $\ket{\filled_2}$ such that $\adj{c_{2,\omega}}\ket{\filled_2} = 0$ $\forall\omega \in (0,\omega\Max)$, the physical occupation in the state $\ket{\vacuum_1,\filled_2}$ is $\sin^2(\theta_{\beta,\mu}(\omega))$ as in \cref{eq:occRel}, so that
\begin{equation}
  \bra{\vacuum_1,\filled_2}
    \adj{c_{1,\omega}} c_{1,\omega}\phantomadj+
    \adj{c_{2,\omega}} c_{2,\omega}\phantomadj
  \ket{\vacuum_1,\filled_2}
  =
  \Tr\bigl(\rho_{\beta,\mu} \adj{f_{1,\omega}}f_{1,\omega}\phantomadj\bigr).
\end{equation}

The interaction Hamiltonian $H\Interaction$ in \cref{eq:intHam} becomes
\begin{multline} \label{eq:intHamThermo}
  H\Interaction =
  \int_{0}^{\omega\Max} \dd\omega\,\bigl[
    h_{\beta,\mu}^1(\omega) (\adj{A\System} c_{1,\omega}\phantomadj + \adj{c_{1,\omega}}A\System\phantomadj)\\
    +h_{\beta,\mu}^2(\omega) (\adj{A\System} c_{2,\omega}\phantomadj + \adj{c_{2,\omega}}A\System\phantomadj)
  \bigr],
\end{multline}
with the coupling $h_{1,2}(\omega)$ defined as
\begin{equation}
  \begin{aligned}
    h_{\beta,\mu}^1(\omega) &= \cos\bigl(\theta_{\beta,\mu}(\omega)\bigr) h(\omega),\\
    h_{\beta,\mu}^2(\omega) &= \sin\bigl(\theta_{\beta,\mu}(\omega)\bigr) h(\omega).
  \end{aligned}
\end{equation}

\section{The Szegő class} \label{sec:szego} \newcommand{\szego}{G}
In this appendix we study the Szegő class of functions to find which conditions we need to impose on the spectral densities so that the modulation and the recombination processes give meaningful results.
It turns out that the conditions are very mild, so that these transformations are always well defined for the usual spectral densities seen in literature.
\subsection{Definitions}
The Szegő class was originally defined in Ref.~\onlinecite[1, Section~12.1]{Szego1975:orthogonal_polynomials} as the set of non-negative measurable functions on \([-\pi,\pi]\) such that
\begin{equation}
  \int_{-\pi}^{\pi}\dd t\,f(t)
  \quad\text{and}\quad
  \int_{-\pi}^{\pi}\dd t\,\abs{\log f(t)}
  \label{szego:original_requirements}
\end{equation}
are finite, or equivalently the set of non-negative measurable functions \(w\) defined on \([-1,1]\) such that \(f(t)\defeq w(\cos t)\abs{\sin t}\) satisfies~\eqref{szego:original_requirements}.

We can rewrite the integrals in~\eqref{szego:original_requirements} to obtain clearer conditions on \(w\).
For the first one, we note that the integrand function is even, so
\begin{equation}
  \int_{-\pi}^{\pi}\dd t\,w(\cos t)\abs{\sin t} = 2\int_{0}^{\pi}\dd t\,w(\cos t)\abs{\sin t};
\end{equation}
then we change variables with \(t=\arccos x\), obtaining
\begin{equation}
  \begin{aligned}
    &2\int_{0}^{\pi}\dd t\,w(\cos t)\abs{\sin t} =\\ &=
    2\int_{-1}^{1}\dd x\,w(x)\abs{\sin(\arccos x)}\frac{1}{\sqrt{1-x^2}}=\\ &=
    2\int_{-1}^{1}\dd x\,w(x)\sqrt{1-x^2}\frac{1}{\sqrt{1-x^2}}=\\ &=
    2\int_{-1}^{1}\dd x\,w(x).
  \end{aligned}
\end{equation}
With the same procedure, the second one becomes instead
\begin{equation}
  \begin{aligned}
    &\int_{-\pi}^{\pi}\dd t\,\abs[\big]{\log\bigl(w(\cos t)\abs{\sin t}\bigr)}=\\ &=
    2\int_{0}^{\pi}\dd t\,\abs[\big]{\log w(\cos t)+\log\abs{\sin t}}=\\ &=
    2\int_{-1}^{1}\dd x\,\frac{\abs{\log w(x)+\log\sqrt{1-x^2}}}{\sqrt{1-x^2}}=\\ &=
    2\int_{-1}^{1}\dd x\,\abs[\bigg]{\frac{\log w(x)}{\sqrt{1-x^2}}+\frac{\log(1-x^2)}{2\sqrt{1-x^2}}}.
  \end{aligned}
\end{equation}
The second fraction in the last integral gives
\begin{equation}
  \int_{-1}^{1}\dd x\,\abs[\bigg]{\frac{\log(1-x^2)}{2\sqrt{1-x^2}}}=-\pi\log 2,
\end{equation}
so on one hand
\begin{equation}
  \begin{aligned}
    &\int_{-1}^{1}\dd x\,\abs[\bigg]{\frac{\log w(x)}{\sqrt{1-x^2}}+\frac{\log(1-x^2)}{2\sqrt{1-x^2}}}\leq\\ &\leq
    \int_{-1}^{1}\dd x\,\abs[\bigg]{\frac{\log w(x)}{\sqrt{1-x^2}}}+\int_{-1}^{1}\dd x\,\abs[\bigg]{\frac{\log(1-x^2)}{2\sqrt{1-x^2}}}=\\ &=
    \int_{-1}^{1}\dd x\,\abs[\bigg]{\frac{\log w(x)}{\sqrt{1-x^2}}}+\pi\log 2
  \end{aligned}
\end{equation}
while with the reverse triangular inequality \(\abs{a-b}\geq\abs[\big]{\abs{a}-\abs{b}}\) we get
\begin{equation}
  \begin{aligned}
    \int\dd x\,\abs{A(x)-B(x)}&\geq
    \int\dd x\,\abs[\big]{\abs{A(x)}-\abs{B(x)}}\geq\\ &\geq
    \abs[\bigg]{\int\dd x\,\bigl(\abs{A(x)}-\abs{B(x)}\bigr)}\geq\\ &\geq
    \int\dd x\,\bigl(\abs{A(x)}-\abs{B(x)}\bigr)=\\ &=
    \int\dd x\,\abs{A(x)}-\int\dd x\,\abs{B(x)}
  \end{aligned}
\end{equation}
so we conclude that
\begin{multline}
  \int_{-1}^{1}\dd x\,\frac{\abs{\log w(x)}}{\sqrt{1-x^2}}-\pi\log 2
  \leq\\ \leq
  \int_{-1}^{1}\dd x\,\abs[\bigg]{\frac{\log w(x)}{\sqrt{1-x^2}}+\frac{\log(1-x^2)}{2\sqrt{1-x^2}}}
  \leq\\ \leq
  \int_{-1}^{1}\dd x\,\frac{\abs{\log w(x)}}{\sqrt{1-x^2}}+\pi\log 2.
\end{multline}
This proves that \(w\) belongs to the Szegő class if and only if
\begin{gather}
  \int_{-1}^{1}\dd x\,w(x)<+\infty,\label{szego:summable_function}
  \\
  \int_{-1}^{1}\dd x\,\frac{\abs{\log w(x)}}{\sqrt{1-x^2}}<+\infty.\label{szego:summable_logarithm}
\end{gather}

Another common definition of the Szegő class is the set of non-negative functions \(w\) on \([-1,1]\) such that
\begin{equation}
  \int_{-1}^{1}\dd x\,\frac{\log w(x)}{\sqrt{1-x^2}}>-\infty
  \label{szego:logarithm_not_minus_infinity}
\end{equation}
which stems from the fact that the asymptotic formula for the recurrence coefficients of the orthogonal polynomials associated to \(w\), found for example in Ref.~\onlinecite[Theorem 12.7.1]{Szego1975:orthogonal_polynomials}, which are key objects in the TEDOPA derivation, contains a factor
\begin{equation}
  \exp\biggl(-\frac{1}{2\pi}\int_{-1}^{1}\dd x\,\frac{\log w(x)}{\sqrt{1-x^2}}\biggr).
\end{equation}
While not always explicitly stated, condition~\eqref{szego:summable_function} should also be required in this case, since the system-environment coupling constant after the TEDOPA transformation is precisely the square root of the left-hand side.

Both definitions imply useful properties, so we start showing that they describe, actually, the same set.
Let \(\szego_1\) be the set of functions satisfying \eqref{szego:summable_function} and \eqref{szego:summable_logarithm}, and \(\szego_2\) the set of those satisfying \eqref{szego:summable_function} and \eqref{szego:logarithm_not_minus_infinity}.
It is easy to see that \(\szego_1\subseteq \szego_2\) since~\eqref{szego:summable_logarithm} bounds the integral in~\eqref{szego:logarithm_not_minus_infinity}.
Vice versa,~\eqref{szego:summable_function} makes the integral in~\eqref{szego:logarithm_not_minus_infinity} also bounded from above: in fact by changing the integration variable with \(x=\cos t\) we first get
\begin{equation}
  \begin{aligned}
    &\int_{-1}^{1}\dd x\,\frac{\log w(x)}{\sqrt{1-x^2}}=\\ &=
    \int_{-\pi}^{0}\dd t\,\frac{\log w(\cos t)}{\sqrt{1-\cos^2t}}\abs{\sin t}=\\ &=
    \int_{-\pi}^{0}\dd t\,\log w(\cos t)=\\ &=
    \int_{-\pi}^{0}\dd t\,\bigl(\log w(\cos t)+\log\abs{\sin t}-\log\abs{\sin t}\bigr)=\\ &=
    \int_{-\pi}^{0}\dd t\,\log \bigl(w(\cos t)\abs{\sin t}\bigr)-
    \int_{-\pi}^{0}\dd t\,\log\abs{\sin t}=\\ &=
    \int_{-\pi}^{0}\dd t\,\log \bigl(w(\cos t)\abs{\sin t}\bigr)+
    \pi\log 2.
  \end{aligned}
\end{equation}
Now we bound the logarithm by its argument, which is of course always positive, and change \(t\) to \(-t\) obtaining
\begin{equation}
  \begin{aligned}
    &\int_{-\pi}^{0}\dd t\,\log \bigl(w(\cos t)\abs{\sin t}\bigr)+\pi\log 2\leq\\ &\leq
    \int_{-\pi}^{0}\dd t\,w(\cos t)\abs{\sin t}+\pi\log 2=\\ &=
    \int_{0}^{\pi}\dd t\,w(\cos t)\abs{\sin t}+\pi\log 2.
  \end{aligned}
\end{equation}
With another change of variables, \(t=\arccos y\), we get to
\begin{equation}
  \begin{aligned}
    &\int_{-1}^{1}\dd y\,w(y)\abs{\sin(\arccos y)}\frac{1}{\sqrt{1-y^2}}+\pi\log 2=\\ &=
    \int_{-1}^{1}\dd y\,w(y)\sqrt{1-y^2}\frac{1}{\sqrt{1-y^2}}+\pi\log 2=\\ &=
    \int_{-1}^{1}\dd y\,w(y)+\pi\log 2
  \end{aligned}
\end{equation}
which we know is finite, so
\begin{equation}
  \int_{-1}^{1}\dd x\,\frac{\log w(x)}{\sqrt{1-x^2}}
  \leq
  \int_{-1}^{1}\dd x\,w(x)+\pi\log 2
  <
  +\infty
\end{equation}
as we wanted.
This proves \eqref{szego:summable_logarithm} and consequently \(\szego_2\subseteq \szego_1\), which ultimately means \(\szego_1=\szego_2\).

We denote the Szegő class just by \(\szego\).
We point out that functions defined on other intervals can be always be translated and dilated so that they become defined in \([-1,1]\); in general, we could as well define the Szegő class on a generic interval \([a,b]\) as the set of functions \(w\colon[a,b]\to[0,+\infty)\) such that
\begin{equation}
  \int_{a}^{b}\dd x\, w(x)<+\infty,
  \quad
  \int_{a}^{b}\dd x\,\frac{\abs{\log w(x)}}{\sqrt{(b-x)(x+a)}}<+\infty
\end{equation}
but it is enough to consider the \([-1,1]\) case, except in cases when we need to add functions defined on different domains, which of course start as Szegő-class functions on different sets.

\subsection{Properties}
Now that we settled on a definition, we analyze what properties we need from Szegő-class functions so that the composition and modulation of spectral densities in the main text is sure to work.
Take for example two functions \(J_1\colon [-1,1]\to[0,+\infty)\) and  \(J_2\colon [0,2]\to[0,+\infty)\), combined into
\begin{equation}
  J'(x)=\Bigl(1+\tanh\frac{x}{2}\Bigr)\bigl(\indicatorf{[-1,1]}(x)J_1(x)+\indicatorf{[0,2]}(x)J_2(x)\bigr),
\end{equation}
which imitates e.g.~\eqref{eq:extTwoVacuum} after ignoring irrelevant constants.
We cannot view this simply as the sum of two functions in the Szegő class on \([-1,1]\) and \([0,2]\) respectively because either i) the functions are on different domains, but functions in \(\szego\) are defined on the \emph{same domain}, or ii) if we consider both of them as defined on \([-1,2]\), then e.g.~\(\indicatorf{[-1,1]}(x)J_1(x)=0\) on \([1,2]\) so cannot be in the Szegő class on \([-1,2]\).
However we can write \(J'\) as
\begin{equation}
  J'(x)=
  \begin{cases}
    (1+\tanh\tfrac{x}{2})J_1(x) & -1\leq x <0,\\
    (1+\tanh\tfrac{x}{2})(J_1(x)+J_2(x)) & 0\leq x <1,\\
    (1+\tanh\tfrac{x}{2})J_2(x) & 1\leq x \leq2\\
  \end{cases}
\end{equation}
which can be seen as the concatenation of the restrictions \(J_1\rvert_{[-1,0]}\), \((J_1+J_2)\rvert_{[0,1]}\) and  \(J_2\rvert_{[1,2]}\), times a bounded function, \(1+\tanh\frac{x}{2}\),  also in the Szegő class.

We need then the Szegő class to be closed under four kinds of operations:
\begin{enumerate}
  \item addition,
  \item product (at least if one of the functions is also bounded almost everywhere),
  \item concatenation,
  \item restriction on a smaller domain.
\end{enumerate}
We study the closure of \(G\) under these operations one by one.
\begin{property}[Addition]
  For any \(f,g\in \szego\),
  \begin{equation}
    \int_{-1}^{1}\dd x\,\bigl(f(x)+g(x)\bigr)=
    \int_{-1}^{1}\dd x\,f(x) + \int_{-1}^{1}\dd x\,g(x)<
    +\infty
  \end{equation}
  and since \(g(x)\geq 0\) for all \(x\in[-1,1]\)
  \begin{equation}
    \int_{-1}^{1}\dd x\,\frac{\log \bigl(f(x)+g(x)\bigr)}{\sqrt{1-x^2}}\geq
    \int_{-1}^{1}\dd x\,\frac{\log f(x)}{\sqrt{1-x^2}}>
    -\infty.
  \end{equation}
\end{property}

\begin{property}[Product with an a.e.\ bounded function] \label{lemma:product_szego}
  Let \(f,g\in \szego\) with \(g\) bounded almost everywhere.
  By Hölder's inequality
  \begin{equation}
    \int_{-1}^{1}\dd x\,f(x)g(x)\leq \lVert f\rVert_1\lVert g\rVert_\infty<+\infty
  \end{equation}
  while for the logarithms
  \begin{multline}
    \int_{-1}^{1}\dd x\,\frac{\log\bigl(f(x)g(x)\bigr)}{\sqrt{1-x^2}}=\\ =
    \int_{-1}^{1}\dd x\,\frac{\log f(x)}{\sqrt{1-x^2}}+
    \int_{-1}^{1}\dd x\,\frac{\log g(x)}{\sqrt{1-x^2}}>
    -\infty
  \end{multline}
  since both are greater than \(-\infty\).
\end{property}
In our case we never multiply two generic spectral densities together: for TCSM we only need to multiply a (shifted) spectral density by factors of the form $\frac{1}{e^x+1}$, $\frac{1}{e^{-x}+1}$ or 
  $1+\tanh\frac{x}{2}$ (with some coefficients), which are always positive and bounded.

\begin{property}[Concatenation] \label{lemma:orthogonal_sum_szego}
  If \(u,v \in \szego\) we define their concatenation as the following function on \([-1,1]\):
  \begin{equation}
    w(t)=
    \begin{cases}
      u(2t+1) & -1\leq t\leq 0,\\
      v(2t-1) & 0<t\leq 1.
    \end{cases}
  \end{equation}
  This function is in \(\szego\) since it obviously satisfies \eqref{szego:summable_function} and
  \begin{equation}
    \begin{aligned}
      &\int_{-1}^{1}\dd t\,\frac{\abs{\log w(t)}}{\sqrt{1-t^2}}=\\ &=
      \int_{-1}^{0}\dd t\,\frac{\abs{\log u(2t+1)}}{\sqrt{1-t^2}}+
      \int_{0}^{1}\dd t\,\frac{\abs{\log v(2t-1)}}{\sqrt{1-t^2}}=\\ &=
      \int_{-1}^{1}\dd x\,\frac{\abs{\log u(x)}}{\sqrt{(3-x)(1+x)}}+
      \int_{-1}^{1}\dd x\,\frac{\abs{\log v(x)}}{\sqrt{(3+x)(1-x)}}\leq\\ &\leq
      \int_{-1}^{1}\dd x\,\frac{\abs{\log u(x)}}{\sqrt{1-x^2}}+
      \int_{-1}^{1}\dd x\,\frac{\abs{\log v(x)}}{\sqrt{1-x^2}}
      <+\infty.
    \end{aligned}
  \end{equation}
\end{property}

\begin{property}[Restriction to subinterval]
  Take \(f(t)\) equal to \(1\) on \([-1,0]\) and \(e^{-1/\sqrt{t}}\) on \((0,1]\): if we restrict it to \([0,1]\) and suitably ``stretch'' it so as to obtain a function on \([-1,1]\), i.e.~\(f'(t):= f(\frac{t+1}{2})\), then \(f'\notin \szego\), so we need more assumptions.
  Informally, we could say that the restriction to a subinterval \([a,1]\) works provided \(0<w(a)<+\infty\) or \(a=-1\) (similarly for a restriction to \([-1,a]\)).
  If \(a=-1\) nothing changes, while if \(a>-1\) and \(0<w(a)<+\infty\) then for a small enough \(\delta>0\)
  \begin{multline}
    \int_{a-\delta}^{a}\dd x\,\frac{\log w(x)}{\sqrt{(a-x)(1+x)}}\sim\\ \sim
    \frac{\log w(a)}{\sqrt{1+a}}\int_{a-\delta}^{a}\dd x\,\frac{1}{\sqrt{a-x}}<+\infty.
  \end{multline}
\end{property}

Note that the assumptions that make this property work are mild (and they are certainly not \emph{necessary}, \(w(a)\) may very well be \(0\) as long as \(w(x)\to 0\) ``nicely'' as \(x\to a\)).

\section{Detailed derivation of FMC} \label{sec:FMC_details}
Here we provide a full derivation of the fermionic Markovian closure discussed in \cref{sec:fermionic_markovian_closure}.

Let us start by considering the TTCF $c_0'(t)$ (see \cref{eq:aux_TTCF}), corresponding to an auxiliary environment starting from the vacuum state. Given the initial condition \(a_j(0)=a_j\), the only non-zero two-time correlation function is
\begin{equation}
  \begin{split}
    c'_0(t)&=
    \vacstate* B\Reservoir(t)\adj{B\Reservoir(0)} \vacstate=\\ &=
    \sum_{k=1}^{N\Closure}\sum_{l=1}^{N\Closure}\zeta_k\conj{\zeta_l}\vacstate* a_k(t)\adj{a_l} \vacstate=\\ &=
    \sum_{k=1}^{N\Closure}\sum_{l=1}^{N\Closure}\zeta_k\conj{\zeta_l}\vacstate* a_k(t)\spstate{l},
  \end{split}
\end{equation}
where \(\spstate{l}\defeq\adj{a_l}\vacstate\) is the state in which the \(l\)-th mode (only) is occupied.

Let \(\vec{v}^i(t)\) be the vector whose \(j\)-th component is \(v^i_j(t)=\vacstate* a_i(t)\spstate{j}\).
It satisfies
\begin{equation}
  \deriv{v^i_j(t)}{t}=
  \vacstate* \dot{a}_i(t) \spstate{j} =
  \vacstate* \adjlindblad\Reservoir\bigl(a_i(t)\bigr) \spstate{j}
\end{equation}
where \(\adjlindblad\Reservoir\) is the adjoint Lindblad operator such that
\begin{equation}
  \adjlindblad\Reservoir(\rho)=
  i\bcomm{H\Reservoir^{(0)},\rho}+\dissipator^{(0)\prime}\Reservoir(\rho)
\end{equation}
which appears in the equation of motion in the Heisenberg picture.
We find
\begin{equation}
  \begin{aligned}
    &\vacstate* H\Reservoir^{(0)} a_i(t) \spstate{j}=
    0,\\
    &\vacstate* a_i(t) H\Reservoir^{(0)} \spstate{j}=
    \sum_{m=1}^{N\Closure}\PMfreqmatemp_{mj} \vacstate* a_i(t) \spstate{m},\\
    &\vacstate* \dissipator\Reservoir^{{(0)}\prime}\bigl(a_i(t)\bigr) \spstate{j}=
    -\frac12 \sum_{m=1}^{N\Closure}\PMdissmatemp_{mj} \vacstate* a_i(t) \spstate{m}.
  \end{aligned}
\end{equation}
We collect these results in the equation
\begin{equation}
  \deriv{\vec{v}^i(t)}{t}=
  \tsp{\bigl(-i\PMfreqmatemp-\tfrac12\PMdissmatemp\bigr)}\vec{v}^i(t),
  \label{eq:eqdiff_funzione_correlazione}
\end{equation}
which together with the initial condition
\begin{equation}
  v^i_j(0)=
  \vacstate* a_i(0) \spstate{j}=
  \vacstate* a_i \spstate{j}=
  \delta_{ij}
\end{equation}
is solved by
\begin{equation}
  \begin{split}
    v^i_j(t)& =
    \sum_{m=1}^{N\Closure} \exp(-it\PMfreqmatemp-\tfrac12t\PMdissmatemp)_{mj}v^i_m(0)=\\
    &= \exp(-it\PMfreqmatemp-\tfrac12t\PMdissmatemp)_{ij}.
  \end{split}
\end{equation}
The correlation function is therefore
\begin{equation}
  \begin{split}
    c'_0(t)&=
    \sum_{k=1}^{N\Closure}\sum_{l=1}^{N\Closure}\zeta_k\conj{\zeta_l}\exp(-it\PMfreqmatemp-\tfrac12t\PMdissmatemp)_{kl}=\\ &=
    \innerp{\zeta, \tsp{\exp(-it\PMfreqmatemp-\tfrac12t\PMdissmatemp)}\zeta}.
  \end{split}
\end{equation}

Now, when the TEDOPA chain starts from the completely filled state \(\filledstate\filledstate*\), we assume
\begin{equation}
  \begin{gathered}
    H\Reservoir^{(1)}=
    \sum_{i,j=1}^{N\Closure}\PMfreqmatfil_{ij}\adj{a_i}a_j,\\
    \dissipator\Reservoir^{(1)}(\rho)=
    \sum_{i,j=1}^{N\Closure}\PMdissmatfil_{ij}\Bigl(\adj{a_i}\rho a_j-\frac12\fcomm{\rho,a_j\adj{a_i}}\Bigr)
  \end{gathered}
\end{equation}
which satisfies \(\dissipator\Reservoir^-(\filledstate\filledstate*)=0\).
The only nonzero correlation function is
\begin{equation}
  \begin{split}
    c'_1(t)&=
    \filledstate* \adj{B\Reservoir(t)} B\Reservoir(0) \filledstate=\\ &=
    \sum_{k=1}^{N\Closure}\sum_{l=1}^{N\Closure}\conj{\zeta_k}\zeta_l\filledstate* \adj{a_k}(t)a_l \filledstate=\\ &=
    \sum_{k=1}^{N\Closure}\sum_{l=1}^{N\Closure}\conj{\zeta_k}\zeta_l\filledstate* \adj{a_k}(t)a_l \filledstate.
  \end{split}
\end{equation}
An analogous calculation brings us to
\begin{equation}
  c'_1(t)=
  \innerp{\zeta, \tsp{\exp(it\PMfreqmatfil-\tfrac12t\PMdissmatfil)}\zeta}.
\end{equation}

The correlation functions \cref{eq:residual_environment_correlation_functions,eq:aux_TTCF} are equal if and only if
\begin{equation}
  \homcoupling^2e^{-i\homfrequency t}C\semicircle(2Kt)=\innerp{\zeta, \tsp{\exp(-it\PMfreqmatemp-\tfrac12t\PMdissmatemp)}\zeta}
  \label{eq:equivalence_condition_empty}
\end{equation}
for the initially empty environment, and
\begin{equation}
  \homcoupling^2e^{i\homfrequency t}C\semicircle(2Kt)=\innerp{\zeta, \tsp{\exp(it\PMfreqmatfil-\tfrac12t\PMdissmatfil)}\zeta}
  \label{eq:equivalence_condition_filled}
\end{equation}
for the initially filled one.
The reduced dynamics will be the same as long as these equalities are satisfied for all \(t \geq 0\).
We note that by taking the conjugate of \cref{eq:equivalence_condition_filled} we have
\begin{equation}
  \begin{split}
    \homcoupling^2e^{-i\homfrequency t}C\semicircle(2Kt)&=
    \conj{\innerp{\zeta, \tsp{\exp(it\PMfreqmatfil-\tfrac12t\PMdissmatfil)}\zeta}}=\\ &=
    \innerp{\zeta, \exp(-it\conj{\PMfreqmatfil}-\tfrac12t\conj{\PMdissmatfil})\zeta}
  \end{split}
  \label{eq:equivalence_condition_filled_conjugated}
\end{equation}
which has the same structure as \cref{eq:equivalence_condition_empty}, so once we solve \cref{eq:equivalence_condition_empty} we can just set \(\PMfreqmatfil=\conj{\PMfreqmatemp}\) and \(\PMdissmatfil=\conj{\PMdissmatemp}\) and obtain a solution for \cref{eq:equivalence_condition_filled}.
Ultimately, this means that we need to solve
\begin{equation}
  \innerp{ \zeta, \tsp{\exp(-it\PMfreqmat-\tfrac12t\PMdissmat)}\,\zeta}=
  \homcoupling^2 e^{-i\homfrequency t}C\semicircle(2\homcoupling t).
  \label{eq:equivalence_condition_generic}
\end{equation}
for \(\PMfreqmat\), \(\PMdissmat\) and \(\zeta\).
We already know from Ref.~\onlinecite{Nuesseler2022:markovian_closure} that a particular solution exists where \(i\PMfreqmat+\frac12\PMdissmat\) is tridiagonal; moreover, we can always rescale a semicircular spectral density so that it becomes the ``unit'' semicircle
\begin{equation}
  j\semicircle(x) = \frac{2}{\pi}\sqrt{1-x^2};
\end{equation}
we can fit a certain set of pseudomodes to reproduce the correlation function generated by \(j\semicircle\), which is simply \(C\semicircle\), and then rescale back the results to obtain the ones relative to the original spectral density.
In detail, let $M$ be an $N\Closure \times N\Closure$ matrix defined as in \cref{eq:mmatrix}  and \(\vec{w}\in\C^{N\Closure}\) be a solution to
\begin{equation}
  \innerp{ w , \exp(tM)w }=
  C\semicircle(t).
\end{equation}
We observe, however, that this equation is the same as the one determined in Ref.~\onlinecite{Nuesseler2022:markovian_closure} for the bosonic case, so the coefficients determined there for the bosonic MC are valid also in our fermionic setting.
The coefficients for the cases $N\Closure = 6,8,10$ are reported in \cref{tab:mc_universal_parameters}.

\begingroup
\setlength{\tabcolsep}{3.5pt}
\newlength{\tableskip}
\setlength{\tableskip}{5ex}
\begin{table}
  \centering
  \begin{filecontents*}{mc_parameters_6.csv}
n,alpha-re,alpha-im,beta-re,beta-im,w-re,w-im
1,-0.0159969178377436,0.0,0.0,0.7852874768986048,2.7407079820293235e-05,-1.114709601324414e-05
2,-1.4774047320945315e-10,0.0,0.0,-0.8125036002430518,-0.4789189669207517,0.3988213902999421
3,-2.175194571396893,0.0,0.0,-1.0787475408258504,6.339100697905341e-06,-3.531951185814045e-06
4,-1.435411352267946e-11,0.0,0.0,-0.6754562463557517,0.4816516807336158,-0.3843678674248487
5,-0.0047911195018862,0.0,0.0,0.8054714129050877,-1.39688879447589e-06,2.446828920026841e-06
6,-1.5738622265225956e-09,0.0,,,0.3828168658192857,-0.2926144102988278
\end{filecontents*}
\pgfplotstabletypeset[
    every head row/.style={
      before row=\toprule,
      after row=\midrule,
    },
    every last row/.style={
    after row=\bottomrule},
    col sep=comma,
    columns={n,alpha-re,beta-im,w-re,w-im},
    columns/n/.style={column name=$j$},
    columns/alpha-re/.style={sci,dec sep align,sci zerofill,precision=2,column name=$\Re\alpha_j$},
    columns/beta-im/.style ={sci,dec sep align,sci zerofill,precision=2,column name=$\Im\beta_j$},
    columns/w-re/.style    ={sci,dec sep align,sci zerofill,precision=2,column name=$\Re w_j$},
    columns/w-im/.style    ={sci,dec sep align,sci zerofill,precision=2,column name=$\Im w_j$},
  ]{mc_parameters_6.csv}

  \vspace{\tableskip}
  \begin{filecontents*}{mc_parameters_8.csv}
n,alpha-re,alpha-im,beta-re,beta-im,w-re,w-im
1,-1.0571454550443759e-09,0.0,0.0,-0.8878551441886222,-0.0658317501607036,-0.2480693025711961
2,-1.6395458268521668e-10,0.0,0.0,0.4072216500650863,-0.1306190039495692,0.0346735860059933
3,-2.696923007439711e-11,0.0,0.0,-0.9963622043839604,-0.1792101501482174,-0.6752581452255852
4,-2.97808855115383,0.0,0.0,-1.4939562087367522,0.0191814455637157,-0.0050848262355202
5,-1.0161736214526636e-09,0.0,0.0,-1.0447733902465357,0.0977165900714343,0.3681497465041551
6,-3.613611453100646e-09,0.0,0.0,-0.4549378571694225,-0.1357210522537042,0.0360130045978403
7,-3.533734455676678e-11,0.0,0.0,0.8475503716040427,-0.1063780972151471,-0.4007719856087376
8,-3.7308241469738736e-11,0.0,,,-0.2912758280346078,0.077306047686776
\end{filecontents*}
\pgfplotstabletypeset[
    every head row/.style={
      before row=\toprule,
      after row=\midrule,
    },
    every last row/.style={
    after row=\bottomrule},
    col sep=comma,
    columns={n,alpha-re,beta-im,w-re,w-im},
    columns/n/.style={column name=$j$},
    columns/alpha-re/.style={sci,dec sep align,sci zerofill,precision=2,column name=$\Re\alpha_j$},
    columns/beta-im/.style ={sci,dec sep align,sci zerofill,precision=2,column name=$\Im\beta_j$},
    columns/w-re/.style    ={sci,dec sep align,sci zerofill,precision=2,column name=$\Re w_j$},
    columns/w-im/.style    ={sci,dec sep align,sci zerofill,precision=2,column name=$\Im w_j$},
  ]{mc_parameters_8.csv}

  \vspace{\tableskip}
  \begin{filecontents*}{mc_parameters_10.csv}
n,alpha-re,alpha-im,beta-re,beta-im,w-re,w-im
1,-0.3430711405230032,0.0,0.0,1.1315098386115716,-0.001318528814401,0.0004616307985947
2,-8.674886121083773e-05,0.0,0.0,1.0453507004042732,0.003320461502283,0.0005489766548683
3,-2.727656090386843,0.0,0.0,-1.0753339704488616,-0.0024046448485234,-0.0014818661614868
4,-0.7093203697466086,0.0,0.0,0.8349475497916327,0.0194192645510923,-0.0355344526619434
5,-3.2354616763218383e-06,0.0,0.0,-0.6039941768050517,-0.0331769757641655,-0.0120047397905429
6,-4.5024755928768747e-07,0.0,0.0,-0.5093377444416272,0.1037507750668207,-0.3530710400477936
7,-2.785321971202375e-06,0.0,0.0,0.6774846403870297,0.1205615272932349,0.0207949826216024
8,-9.476253878456016e-05,0.0,0.0,0.1605800966525998,0.1647842706995822,-0.8174053022264767
9,-0.0013707004912771,0.0,0.0,-0.9507225900537972,-0.1211749374501335,-0.0044462242353818
10,-5.954027014131764e-06,0.0,,,0.0471863161980153,-0.3668399977306861
\end{filecontents*}
\pgfplotstabletypeset[
    every head row/.style={
      before row=\toprule,
      after row=\midrule,
    },
    every last row/.style={
    after row=\bottomrule},
    col sep=comma,
    columns={n,alpha-re,beta-im,w-re,w-im},
    columns/n/.style={column name=$j$},
    columns/alpha-re/.style={sci,dec sep align,sci zerofill,precision=2,column name=$\Re\alpha_j$},
    columns/beta-im/.style ={sci,dec sep align,sci zerofill,precision=2,column name=$\Im\beta_j$},
    columns/w-re/.style    ={sci,dec sep align,sci zerofill,precision=2,column name=$\Re w_j$},
    columns/w-im/.style    ={sci,dec sep align,sci zerofill,precision=2,column name=$\Im w_j$},
  ]{mc_parameters_10.csv}
  \caption{%
    Universal parameters for the Markovian closure in the \(N\Closure=6,8\) and \(10\) cases (from top to bottom).
    In all three cases \(\Im\alpha_j\) and \(\Re\beta_j\) are zero for all \(j\).
  }
  \label{tab:mc_universal_parameters}
\end{table}
\endgroup

Going back to \cref{eq:equivalence_condition_generic}, define \(\PMfreqmat'\defeq\PMfreqmat-\homfrequency I\) and \(\vec{\zeta}'\defeq\homcoupling\vec{\zeta}\): we get
\begin{align*}
  \homcoupling^2 e^{-i\homfrequency t}\innerp{ \zeta' , \tsp{\exp(-it\PMfreqmat'-\tfrac12t\PMdissmat)}\,\zeta' }&=
  \homcoupling^2 e^{-i\homfrequency t}C\semicircle(2\homcoupling t)\\
  \innerp{ \zeta' , \tsp{\exp(-it\PMfreqmat'-\tfrac12t\PMdissmat)}\,\zeta' }&=
  C\semicircle(2\homcoupling t);\\
  \intertext{now rescale the matrices by \(2\homcoupling\), with \(\PMfreqmat'=2\homcoupling\PMfreqmat''\) and \(\PMdissmatfil=2\homcoupling\PMdissmat''\) to obtain}
  \innerp{ \zeta' , \tsp{\exp(-2i\homcoupling t\PMfreqmat''-\tfrac122\homcoupling t\PMdissmat'')}\,\zeta' }&=
  C\semicircle(2\homcoupling t),\\
  \intertext{then rescale the time, too, with \(s=2\homcoupling t\), so that}
  \innerp{ \zeta' , \tsp{\exp(-is\PMfreqmat''-\tfrac12s\PMdissmat'')}\,\zeta' }&=
  C\semicircle(s).
\end{align*}
This shows that \(M=-i\PMfreqmat''-\tfrac12\PMdissmat''\) and \(\vec w=\vec{\zeta}'\); therefore, if we set $\PMfreqmatemp$ as in \cref{eq:lambdazero} and \(\PMdissmat_{jk}=\delta_{jk}\gamma_j\) (see \cref{eq:gammazero}), we obtain the following relations:
\begin{equation}
  \begin{aligned}
    \omega_j &= \homfrequency-2\homcoupling\Im\alpha_j,\\
    \gamma_j &= -4\homcoupling\Re\alpha_j,\\
    g_j      &= -2\homcoupling\Im\beta_j,\\
    \zeta_j  &= \homcoupling w_j.
  \end{aligned}
\end{equation}
Recalling \cref{eq:equivalence_condition_filled_conjugated}, if the initial state is the filled one we will have to use \(\conj{\alpha_j}\) and \(\conj{\beta_j}\) instead of \(\alpha_j\) and \(\beta_j\) respectively.

\section{Convergence of chain coefficients from simulations}
\label{sec:convergence_chain_coefficients_from_simulations}
In this section we shortly comment on the convergence of the chain coefficients of the equivalent environments from the non-interacting SIAM simulations in \cref{fig:benchmark_spinless_siam_mu1,fig:benchmark_spinless_siam_mu0.2_NE8}, which are described by a semicircle spectral density function on \((0,2)\) with \(\mu=1\) and \(\mu=\num{0.2}\) respectively, at temperature \(T=\num{0.4}\).
It is clear from the plots in \cref{fig:correlation_function_check_semicircle_mu1,fig:correlation_function_check_semicircle_mu0.2} that the chain coefficients are very close to their asymptotic values already from \(n=5\), which motivates the choices for \(N\Environment\) in the simulations shown in the main text.

\begin{figure}
  \centering
  \begingroup

\pgfplotstableread[col sep=comma]{data_semicircle_T4_mu1.thermofield}\coefficients
\pgfplotstableread[col sep=comma, x=frequency]{data_correlation_function_check_semicircle_mu1.csv}\data

\pgfmathsetmacro{\vsep}{1.5 cm}
\pgfmathsetmacro{\hsep}{0.6 cm}
\pgfmathsetmacro{\pwidth}{\columnwidth-0.5cm}
\pgfmathsetmacro{\pheight}{\columnwidth-0.5cm}

\pgfdeclarelayer{background}
\pgfdeclarelayer{foreground}
\pgfsetlayers{background,main,foreground}

\begin{tikzpicture}[remember picture]
  \begin{pgfonlayer}{background}
    \begin{groupplot}[
        group style = {
          rows = 2,
          columns = 1,
          vertical sep = \vsep,
        },
        xtick align = outside,
        ytick align = outside,
        ylabel near ticks,
        width = \pwidth,
        grid = major,
        no markers, 
        every axis plot/.append style={very thick},
      ]
      \nextgroupplot[
        height=0.75*\pheight,
        table/x expr=\coordindex,
        xmax=5,
        ymode=log,
        xlabel=$n$,
        legend columns = 1,
        legend style={
          at={(0.05,0.05)},
          anchor=south west,
          /tikz/every even column/.append style={column sep=0.5cm},
          draw=none,
          font=\small
        }
      ]
      \addplot+ [solid, strongred]   table [y expr=abs(1/2-\thisrow{coupempty})]  \coefficients;
      \addplot+ [solid, strongblue]  table [y expr=abs(1/2-\thisrow{coupfilled})] \coefficients;
      \addplot+ [dotted, strongred]  table [y expr=abs(0-\thisrow{freqempty})]    \coefficients;
      \addplot+ [dotted, strongblue] table [y expr=abs(0-\thisrow{freqfilled})]   \coefficients;
      %
      \legend{
        {$\abs{\kappa_{0,n}-\homcoupling}$},
        {$\abs{\kappa_{1,n}-\homcoupling}$},
        {$\abs{\omega_{0,n}-\homfrequency}$},
        {$\abs{\omega_{1,n}-\homfrequency}$},
      }
      \nextgroupplot[
        height = \pheight,
        xlabel = $\omega$,
        xmin = -1.5,
        xmax = 1.5,
        ymin = -0.007,
        ymax = 0.06,
      ]
      \addplot+ [strongred!50] table [y expr=(0.15811)^2/(2*pi)*\thisrow{empty6simulated}] \data
      node [strongred, pos=0.65, above] {\(J^{(0)}(\omega)\)};
      \addplot+ [black, dashed] table [y expr=\thisrow{emptyexpected}] \data;
      \addplot+ [strongblue!50] table [y expr=(0.15811)^2/(2*pi)*\thisrow{filled6simulated}] \data
      node [strongblue, pos=0.35, above] {\(J^{(1)}(\omega)\)};
      \addplot+ [black, dashed] table [y expr=\thisrow{filledexpected}] \data;
      \coordinate (inset1) at (rel axis cs: 0.95,0.95);
    \end{groupplot}
  \end{pgfonlayer}
  \begin{pgfonlayer}{foreground}
    \begin{semilogyaxis}[
        at={(inset1)},
        name = insetaxis,
        axis background/.style={fill=white},
        anchor={outer north east},
        footnotesize,
        xtick align = outside,
        ytick align = outside,
        ylabel near ticks,
        height = 0.4*\pheight,
        width = 0.4*\pwidth+\hsep,
        xmin = -1.5,
        xmax= 1.5,
        grid = major,
        no markers,
        table/x = frequency, 
      ]
      \addplot+ [strongred] table [y expr=abs(\thisrow{emptyexpected}-(0.15811)^2/(2*pi)*\thisrow{empty6simulated})] \data;
      \addplot+ [strongblue] table [y expr=abs(\thisrow{filledexpected}-(0.15811)^2/(2*pi)*\thisrow{filled6simulated})] \data;
    \end{semilogyaxis}
  \end{pgfonlayer}
  \begin{pgfonlayer}{main}
    \fill [black!0] ([shift={(-1pt,-1pt)}] insetaxis.outer south west)
      rectangle ([shift={(+5pt,+5pt)}] insetaxis.outer north east);
  \end{pgfonlayer}
\end{tikzpicture}
\endgroup
  \caption{%
    Absolute distance between chain coefficients of the equivalent environments (\(j=0\): initially empty, \(j=1\): initially filled) of the simulations in \cref{fig:benchmark_spinless_siam_mu1}.
    Note that in the top plot, since \(\mu=1\) lies in the middle of the support of the original spectral density function, the equivalent environments are such that \(J^{(0)}(\omega)=J^{(1)}(-\omega)\), which results in \(\Omega=0\), \(\omega_{0,n}=-\omega_{1,n}\), and \(\kappa_{0,n}=\kappa_{1,n}\) for all \(n\), hence the red and blue lines in the top plot coincide.
  }
  \label{fig:correlation_function_check_semicircle_mu1}
\end{figure}

\begin{figure}
  \centering
  \begingroup

\pgfplotstableread[col sep=comma]{data_semicircle_T4_mu0.2.thermofield}\coefficients
\pgfplotstableread[col sep=comma, x=frequency]{data_correlation_function_check_semicircle_mu0.2.csv}\data

\pgfmathsetmacro{\vsep}{1.5 cm}
\pgfmathsetmacro{\hsep}{0.6 cm}
\pgfmathsetmacro{\pwidth}{\columnwidth-0.5cm}
\pgfmathsetmacro{\pheight}{\columnwidth-0.5cm}

\pgfdeclarelayer{background}
\pgfdeclarelayer{foreground}
\pgfsetlayers{background,main,foreground}

\begin{tikzpicture}[remember picture]
  \begin{pgfonlayer}{background}
    \begin{groupplot}[
        group style = {
          rows = 2,
          columns = 1,
          vertical sep = \vsep,
        },
        xtick align = outside,
        ytick align = outside,
        ylabel near ticks,
        width = \pwidth,
        grid = major,
        no markers, 
        every axis plot/.append style={very thick},
      ]
      \nextgroupplot[
        height=0.75*\pheight,
        table/x expr=\coordindex,
        xmax=5,
        ymode=log,
        xlabel=$n$,
        legend columns = 1,
        legend style={
          at={(0.05,0.05)},
          anchor=south west,
          /tikz/every even column/.append style={column sep=0.5cm},
          draw=none,
          font=\small
        },
        every axis plot/.append style={very thick},
      ]
      \addplot+ [solid, strongred]   table [y expr=abs(1/2-\thisrow{coupempty})]  \coefficients;
      \addplot+ [solid, strongblue]  table [y expr=abs(1/2-\thisrow{coupfilled})] \coefficients;
      \addplot+ [dotted, strongred]  table [y expr=abs(8/10-\thisrow{freqempty})]    \coefficients;
      \addplot+ [dotted, strongblue] table [y expr=abs(8/10-\thisrow{freqfilled})]   \coefficients;
      \legend{
        {$\abs{\kappa_{0,n}-\homcoupling}$},
        {$\abs{\kappa_{1,n}-\homcoupling}$},
        {$\abs{\omega_{0,n}-\homfrequency}$},
        {$\abs{\omega_{1,n}-\homfrequency}$},
      }
      \nextgroupplot[
        height = \pheight,
        xlabel = $\omega$,
        xmin = -0.4,
        xmax = 2,
        ymin = -0.01,
        ymax = 0.08,
        every axis plot/.append style={very thick},
      ]
      \addplot+ [strongred!50] table [y expr=(0.203)^2/(2*pi)*\thisrow{empty8simulated}] \data
      node [strongred, pos=0.6, above right] {\(J^{(0)}(\omega)\)};
      \addplot+ [black, dashed] table [y expr=\thisrow{emptyexpected}] \data;
      \addplot+ [strongblue!50] table [y expr=(0.0937)^2/(2*pi)*\thisrow{filled8simulated}] \data
      node [strongblue, pos=0.4, above right] {\(J^{(1)}(\omega)\)};
      \addplot+ [black, dashed] table [y expr=\thisrow{filledexpected}] \data;
      \coordinate (inset1) at (rel axis cs: 0.95,0.95);
    \end{groupplot}
  \end{pgfonlayer}
  \begin{pgfonlayer}{foreground}
    \begin{semilogyaxis}[
        at={(inset1)},
        name = insetaxis,
        axis background/.style={fill=white},
        anchor={outer north east},
        footnotesize,
        xtick align = outside,
        ytick align = outside,
        ylabel near ticks,
        height = 0.4*\pheight,
        width = 0.4*\pwidth+\hsep,
        xmin = -0.4,
        xmax= 2,
        grid = major,
        no markers,
        table/x = frequency, 
      ]
      \addplot+ [strongred] table [y expr=abs(\thisrow{emptyexpected}-(0.203)^2/(2*pi)*\thisrow{empty8simulated})] \data;
      \addplot+ [strongblue] table [y expr=abs(\thisrow{filledexpected}-(0.0937)^2/(2*pi)*\thisrow{filled8simulated})] \data;
    \end{semilogyaxis}
  \end{pgfonlayer}
  \begin{pgfonlayer}{main}
    \fill [black!0] ([shift={(-1pt,-1pt)}] insetaxis.outer south west)
      rectangle ([shift={(+5pt,+5pt)}] insetaxis.outer north east);
  \end{pgfonlayer}
\end{tikzpicture}
\endgroup
  \caption{%
    Absolute distance between chain coefficients of the equivalent environments (\(j=0\): initially empty, \(j=1\): initially filled) of the simulations in \cref{fig:benchmark_spinless_siam_mu0.2_NE8} (bottom).
  }
  \label{fig:correlation_function_check_semicircle_mu0.2}
\end{figure}

%

\end{document}